\newcommand{\Zcal}{\mathcal{Z}}
\newcommand{\signeff}{s_{\mathrm{eff}}}

\documentclass{aa-fixed-texlive2022}

\usepackage{graphicx}
\usepackage{txfonts}

\usepackage[colorlinks,allcolors=blue]{hyperref}

\usepackage{mathtools}

\begin{document}

   \title{Aether scalar tensor theory confronted with weak lensing data at small accelerations}

  \author{T. Mistele
          \inst{1,2}
          \and
          S. McGaugh\inst{2}
          \and
          S. Hossenfelder \inst{1,3}
          }

   \institute{Frankfurt Institute for Advanced Studies,
              Ruth-Moufang-Str. 1, D 60438 Frankfurt am Main, Germany\\
              \email{mistele@fias.uni-frankfurt.de}
         \and
             Department of Astronomy, Case Western Reserve University, 10900 Euclid Avenue,
             Cleveland, OH 44106, USA
         \and
         Munich Center for Mathematical Philosophy, Ludwig-Maximilians-Universität, Geschwister-Scholl-Platz 1, D-80539 München, Germany\\
             }

   \date{\today}

  \abstract
   {The recently proposed aether scalar tensor (AeST) model  reproduces both the successes of particle dark matter on cosmological scales and those of modified Newtonian dynamics (\textsc{MOND}) on galactic scales. But the AeST model reproduces \textsc{MOND} only up to a certain maximum galactocentric radius. Since \textsc{MOND} is known to fit very well to observations at these scales, this raises the question of whether the AeST model comes into tension with data.}
    {We tested whether or not the AeST model is in conflict with observations using a recent analysis of data for weak gravitational lensing.}
   {We solved the equations of motion of the AeST model, analyzed the solutions' behavior, and compared the results to observational data.}
   {The AeST model shows some deviations from \textsc{MOND} at the radii probed by weak gravitational lensing. 
   The data show no clear indication of these predicted deviations.}
   {}


   \maketitle

\section{Introduction}
\label{sec:introduction}

Recently, various models have been proposed that combine the successes of modified Newtonian dynamics \cite[MOND,][]{Milgrom1983a, Milgrom1983b, Milgrom1983c, Bekenstein1984} on galactic scales with those of the $\Lambda$ cold dark matter model (\textsc{$\Lambda$CDM}) on cosmological scales.
Examples are superfluid dark matter \cite[SFDM,][]{Berezhiani2015, Berezhiani2018}, the aether scalar tensor (AeST) model \citep{Skordis2020, Skordis2021}, and the neutrino-based model by \citet{Angus2009} ($\nu$HDM).
The focus of the present paper is on the AeST model. An accompanying paper will look at \textsc{SFDM}.

The AeST model is a relativistic model that can reproduce \textsc{MOND} in the vicinity of galaxies and fits the fluctuations in the cosmic microwave background (CMB) as well as the matter power spectrum.
In addition, it has a tensor mode that propagates at the speed of light which avoids difficulties matching the observations associated with GW170817 \citep{Sanders2018, Boran2018}.

An important ingredient in the AeST model is a so-called ghost condensate \citep{Arkani-Hamed2004, Arkani-Hamed2007}.
This ghost condensate is the major difference between the action of the AeST model in the static limit and the standard MOND-type action for multifield theories \citep{Famaey2012}.
The ghost condensate has an energy density that acts as an additional source for the gravitational field equations.

The integrated mass of the ghost condensate is generally negligible close to galaxies, where rotation curves are measured. Beyond a few hundred kiloparsecs, however, the ghost condensate mass is no longer negligible compared to the baryonic mass which leads to deviations from \textsc{MOND}.

This is a desired feature for galaxy clusters where observations require accelerations larger than what \textsc{MOND} predicts \citep{Aguirre2001, Sanders2003, Eckert2022}.
It may, however, be in conflict with unprecedented recent observations at  large radii around galaxies: The analysis of weak-gravitational lensing data from \citet{Brouwer2021} found \textsc{MOND}-like behavior around galaxies up to $\sim1\,\mathrm{Mpc}$.
Here, we explore whether this finding is compatible with the AeST model.

Since MOND is known to fit these observations well, we adopt an indirect approach to comparing the AeST model to observations.
First, we introduce a method to quantify the deviation of the AeST model from MOND and analyze for which solutions these deviations are minimal.
We then compare these optimal solutions -- as well as slightly suboptimal ones -- to the observational data.

\section{Equations of motion and chemical potential}

For galaxies we can use the quasi-static weak-field limit of the AeST model.
In this limit, the model can be described by two fields, $\hat{\Phi}$ and $\varphi$, whose equations of motion are (at least in the spherically symmetric case that interests us here, see Appendix~\ref{sec:appendix:actionandeom}),
\begin{subequations}
\label{eq:eom}
\begin{align}
\Delta \hat{\Phi} &= f_G \cdot 4 \pi G_N \left(\rho_b + \rho_c\right) \,, \\
\vec{\nabla} \left( \tilde{\mu}\left(\frac{|\vec{\nabla} \varphi|}{a_0}\right) \vec{\nabla} \varphi \right) &=  f_G \cdot 4 \pi G_N \left(\rho_b + \rho_c \right) \,,
\end{align}
\end{subequations}
where $a_0$ is the MOND acceleration scale and $f_G$ is the conversion factor between Newton's gravitational constant $G_N$ and the constant $\hat{G}$ that appears in the Lagrangian (see Appendix~\ref{sec:appendix:actionandeom}).
We use $a_0 = 1.2 \cdot 10^{-10}\,\mathrm{m}/\mathrm{s}^2$ \citep{Lelli2017b}.
Both fields are sourced by the baryonic energy density $\rho_b$ and the ghost condensate density $\rho_c$.
The function $\tilde{\mu}$ can be freely chosen in this model and corresponds to an interpolation function of MOND.
That is, it determines how the model interpolates between Newtonian gravity at large accelerations and MOND-like gravity at small accelerations.
The constant $f_G$ determines how much of the total acceleration (see below) in the Newtonian limit is due to $\varphi$ and how much is due to $\hat{\Phi}$.

In the AeST model, ordinary matter couples to the metric $g_{\mu \nu}$ in the usual way.
The metric has the same form as in general relativity (GR), but with the Newtonian potential $\Phi$ being a combination of $\hat{\Phi}$ and $\varphi$, namely $\Phi \equiv \hat{\Phi} + \varphi$.
Thus, the total acceleration felt by matter is $\vec{a}_{\mathrm{tot}} \equiv \vec{a}_{\hat{\Phi}} + \vec{a}_{\varphi} \equiv - \vec{\nabla} \hat{\Phi} - \vec{\nabla} \varphi $.

The ghost condensate energy density $\rho_c$ is given by
\begin{align}
 \label{eq:rhoc}
 \rho_c = \frac{m^2}{4 \pi G_N f_G} \left(\frac{\dot{\varphi}}{Q_0} -\hat{\Phi} - \varphi\right) \,,
\end{align}
where $m$ and $Q_0$ are constants.
For cosmology, \citet{Skordis2020} considered nonlinear corrections to this.
We do not include these here for reasons discussed in Appendix~\ref{sec:appendix:higherorder}.

We did not set $\dot{\varphi} = 0$ in Eq.~\eqref{eq:rhoc} because any constant $\dot{\varphi} = - \dot{\hat{\Phi}}$ still gives time-independent equations of motion.
Indeed, $\dot{\varphi}$ represents the chemical potential of the condensate.
To see this, one first observes that the model is shift-symmetric under $\varphi \to \varphi + \tilde{c}$, $\hat{\Phi} \to \hat{\Phi} - \tilde{c}$ for any constant $\tilde{c}$.
In general, to describe equilibrium states, one introduces a chemical potential $\mu$ for each symmetry by shifting the Hamiltonian $H$ by $H \to H - \mu \, Q$ where $Q$ is the conserved quantity associated with the symmetry.
In the AeST model and on the level of the Lagrangian, this corresponds to shifting $\dot{\varphi} \to \dot{\varphi} + \mu$ and $\dot{\hat{\Phi}} \to \dot{\hat{\Phi}} - \mu$ \citep{Mistele2019, Kapusta1981, Haber1982, Bilic2008} or equivalently to considering solutions with $\dot{\varphi} = \mu$ and $\dot{\hat{\Phi}} = - \mu$. (We note that the parameter $m$ was called $\mu$ in \citet{Skordis2020}. We use $\mu$ instead to denote the chemical potential.)

Consequently, the behavior of the AeST model around galaxies depends on the choice of this chemical potential of the ghost condensate.
This corresponds to a choice of boundary condition for the combination $\mu/Q_0 - \Phi$ which is a gauge-invariant variable as shown in \citet{Skordis2021}.
This is an important difference to \textsc{MOND} where $\rho_b$ alone determines the phenomenology around galaxies.

For real galaxies, these chemical potentials are ultimately set by galaxy formation.
Since no simulations of nonlinear structure formation in the AeST model are available,
  we treated the chemical potential of each galaxy as a free parameter.
In this regard, the AeST model is similar to SFDM.
Both models require a choice of chemical potential to make predictions in galaxies \citep{Berezhiani2015, Berezhiani2018, Mistele2019, Hossenfelder2020, Mistele2020, Mistele2022}.

We now assume spherical symmetry.
Then, solutions have the same form as in MOND, just with the baryonic mass replaced by an effective mass that includes the condensate mass,
\begin{align}
 M_{\mathrm{eff}}(r) \equiv M_b(r) + M_c(r) \equiv 4\pi \int_0^r dr' r'^2 (\rho_b(r') +  \rho_c(r')) \,.
\end{align}
Thus, with the Newtonian acceleration $a_N$ in the negative radial direction,
\begin{align}
 a_N(r) \equiv a_b(r) + a_c(r) \equiv \frac{G_N M_b(r)}{r^2} + \frac{G_N M_c(r)}{r^2} \,,
\end{align}
we can write the total acceleration in the negative radial direction $a_{\mathrm{tot}}$ as
 $a_{\mathrm{tot}} = a_N(r) \cdot \nu\left(|a_N(r)|/a_0\right)$
where the interpolation function $\nu$ is determined by the free function $\tilde{\mu}$ in the AeST model (see Eq.~\eqref{eq:eom}).
Unlike in MOND, $a_N$ can be negative because the condensate mass $M_c$ can be negative.
We discuss this in more detail below.

For simplicity, we chose the interpolation function such that
\begin{align}
 a_{\mathrm{tot}}(r) = \signeff \cdot \left(|a_N(r)| + \sqrt{a_0 |a_N(r)|} \right) \,,
\end{align}
where $\signeff$ is the sign of $M_{\mathrm{eff}}$.
We recover MOND by leaving out the contributions from the condensate, $M_c = 0$.
Thus, when comparing the AeST model to MOND below, we assumed the following acceleration for MOND,
\begin{align}
a_{\mathrm{MOND}}(r) \equiv a_b(r) + \sqrt{a_0 a_b(r)}\,.
\end{align}
This assumes the same interpolation function for MOND and the AeST model.

This interpolation function has the correct limits, namely $a_{\mathrm{MOND}} \to a_b$ for accelerations much larger than $a_0$ and $a_{\mathrm{MOND}} \to \sqrt{a_0 a_b}$ for accelerations much smaller than $a_0$.
Still, in general, this choice is too simplistic.
But it suffices for our purposes because we are only interested in the small-acceleration regime where all interpolation functions give $a_{\mathrm{MOND}} \approx \sqrt{a_0 a_b}$.

The total acceleration $a_{\mathrm{tot}}$ in the AeST model does not explicitly depend on $f_G$ and is the sum of $a_{\hat{\Phi}} = f_G \, a_N$ and $a_{\varphi} =\signeff ( (1-f_G) \, |a_N| + \sqrt{a_0 |a_N|})$.
Below, we refer to its MOND-like part $\sqrt{a_0 |a_N|}$ and its Newton-like part $a_N$, respectively, as
\begin{align}
 a_{\tilde{\Phi}} \equiv \signeff |a_N| \,, \quad a_{\tilde{\varphi}} \equiv \signeff \sqrt{a_0 |a_N|} \,.
\end{align}

In the following we assumed point particle baryonic masses for simplicity.
This suffices for our purposes, because the details of the baryonic mass distribution do not matter much at the large galactocentric radii we consider.
See Appendix~\ref{sec:approximation} for an approximate analytical solution for this case and Appendix~\ref{sec:appendix:numerical} for how we calculate numerical solutions.

\section{Deviations from MOND}

\begin{figure}
 \centering
 \includegraphics[width=\hsize]{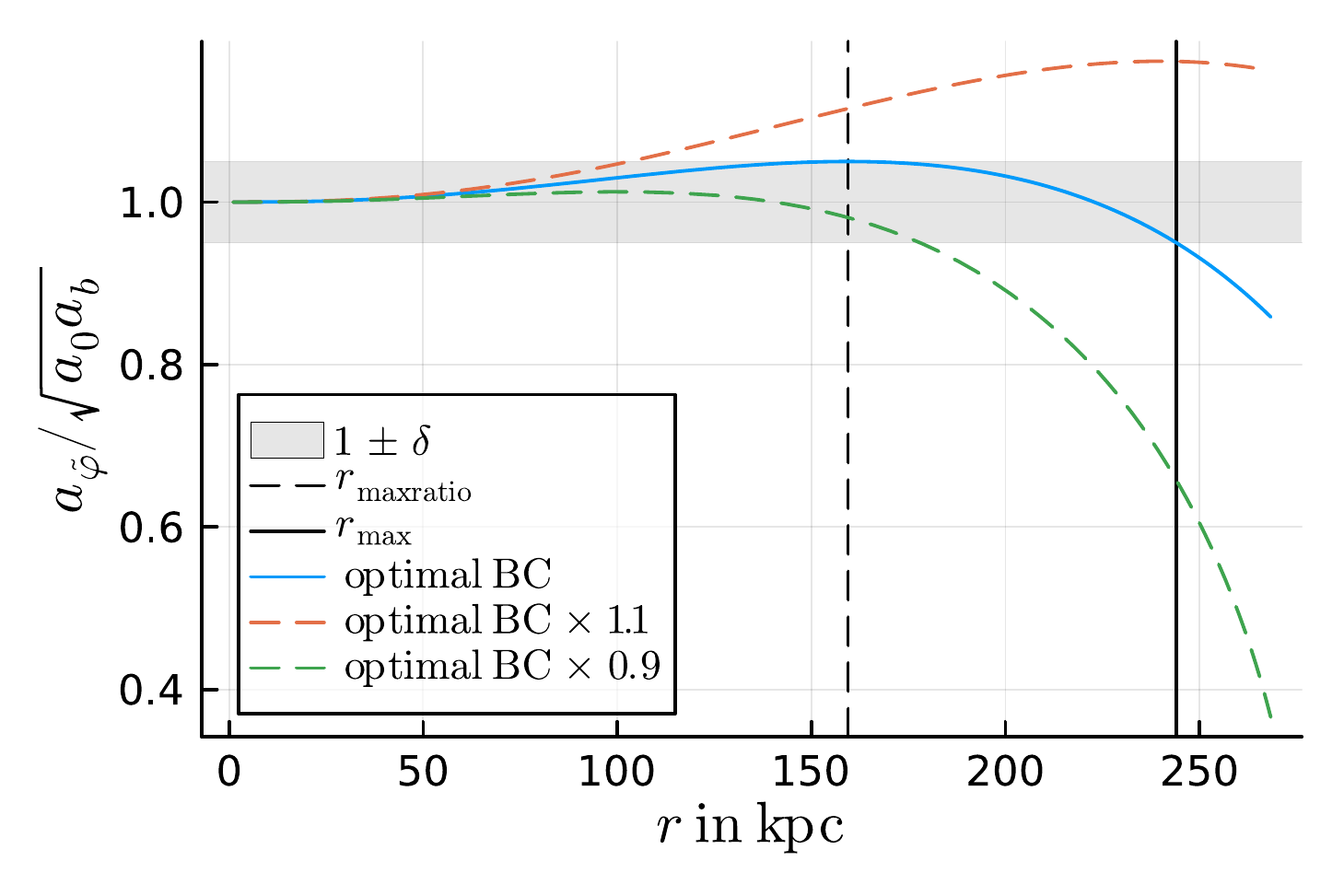}
 \caption{
   Numerical solution with the optimal boundary condition for $\delta = 0.05$ (solid blue line).
   That is, the solution for which $a_{\tilde{\varphi}}$ stays within a fraction $\delta$ of the MOND-like acceleration $\sqrt{a_0 a_b}$ up to the maximum possible radius $r_{\mathrm{max}}$.
   This is for $M_b = 2 \cdot 10^{10}\,M_\odot$ and $f_G/m^2 = 0.99\,\mathrm{Mpc}^2$.
   The dashed red and green lines show solutions with the optimal boundary condition multiplied by $1.1$ and $0.9$, respectively.
   Vertical black lines indicate the maximum radius $r_{\mathrm{max}}$ and the radius where the optimal solution reaches its maximum $r_{\mathrm{maxratio}}$.
 }
 \label{fig:illustrate-bestIC-terminology}
\end{figure}

The AeST model reproduces MOND so long as the condensate's total mass is small compared to the baryonic mass.
This condition is usually fulfilled in the inner parts of galaxies but not far away from the galaxy.
Indeed,  given a maximum allowed fractional deviation $\delta$ from MOND,
  there is an optimal boundary condition for which the MOND-like behavior extends to a finite maximum radius $r_{\mathrm{max}}$.
For all other boundary conditions, deviations from MOND set in earlier.
This is illustrated in Fig.~\ref{fig:illustrate-bestIC-terminology}.

Specifically, we imposed the maximum allowed deviation $\delta$ as
\begin{align}
\label{eq:deltadef}
\left|\frac{a_{\tilde{\varphi}}(r)}{\sqrt{a_0 a_b(r)}} - 1 \right| = \left|\frac{\signeff \cdot \sqrt{a_0 |a_N(r)|}}{\sqrt{a_0 a_b(r)}} - 1 \right| \leq \delta \,.
\end{align}
That is, we compared the accelerations in the AeST model and MOND, focusing on the MOND-like contributions $\sqrt{a_0 a_N}$ and $\sqrt{a_0 a_b}$.
Alternatively, one could compare the total acceleration in both models, that is, $a_N + \sqrt{a_0 a_N}$ and $a_b + \sqrt{a_0 a_b}$.
But at the large radii we consider here, the difference is negligible (see Appendix~\ref{sec:optimal}) and we used Eq.~\eqref{eq:deltadef} for simplicity.

The maximal radius $r_{\mathrm{max}}$, up to which we allowed accelerations to deviate by less than a fraction $\delta$ from MOND, is given by
\begin{align}
 \label{eq:rmax}
 \frac{r_{\mathrm{max}}}{r_{\mathrm{MOND}}} \approx 1.53\, \left(9 \frac{(1+\delta)^2-1}{r_{\mathrm{MOND}}^2 \,m^2/f_G}\right)^{1/3}  \,.
\end{align}
Here, $r_{\mathrm{MOND}}$ is a constant known as the MOND-radius $\sqrt{G_N M_b/a_0}$.
See Appendix~\ref{sec:optimal} for a derivation of those properties and an analysis of the conditions under which this estimate holds.

\begin{figure}
 \centering
 \includegraphics[width=\hsize]{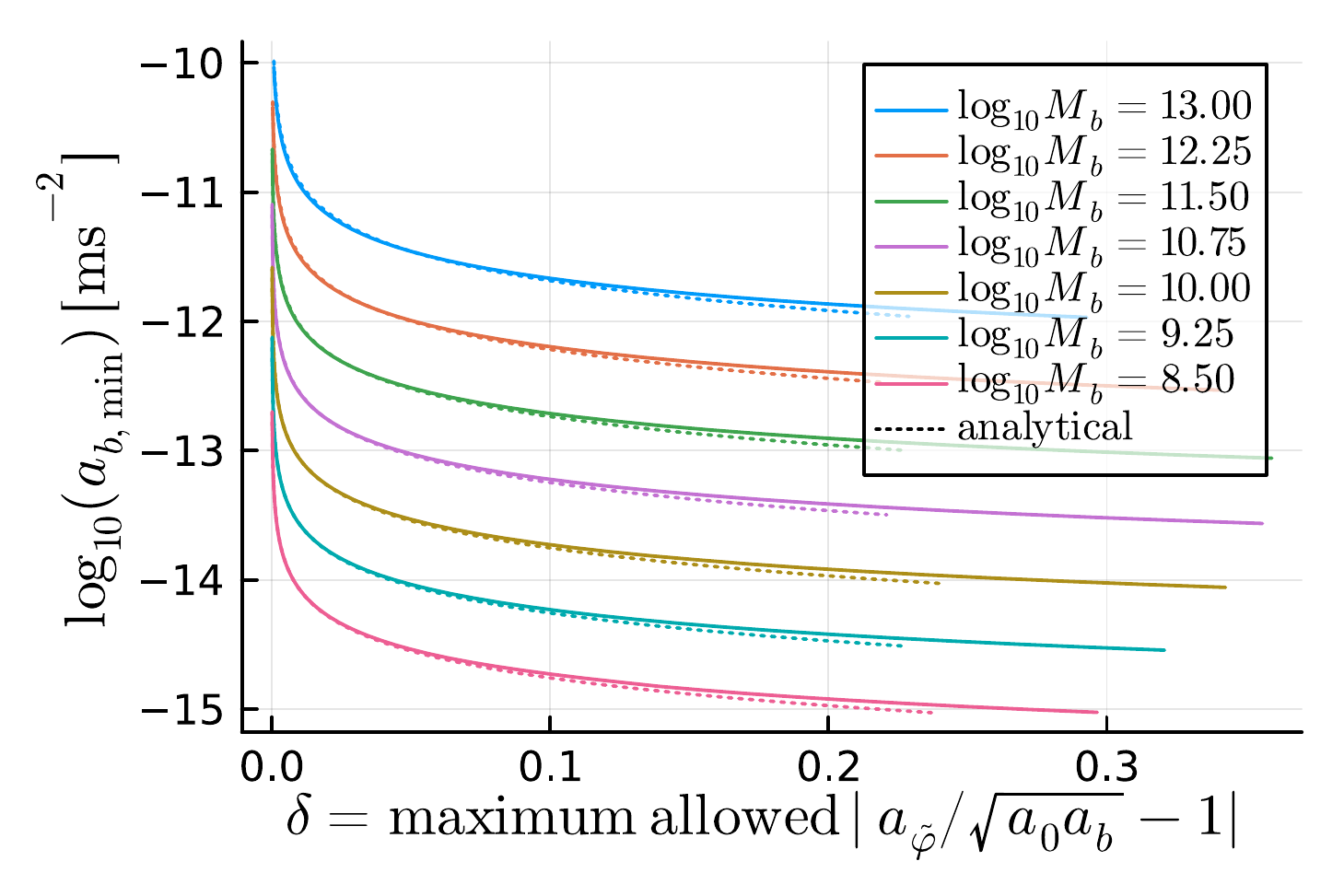}
 \caption{
   Acceleration $a_{b,\mathrm{min}} = G_N M_b/r_{\mathrm{max}}^2$ down to which the acceleration $a_{\tilde{\varphi}} = \signeff \sqrt{a_0 |a_N|}$ can, at best, stay within a fraction $\delta$ of the MOND-like acceleration $\sqrt{a_0 a_b}$ as a function of $\delta$.
   This is for $f_G/m^2 = 0.99\,\mathrm{Mpc}^2$.
   We show the result for both numerical (solid lines) and analytical (dotted lines) solutions and for various baryonic masses $M_b$.
   The analytical approximation is shown only where our estimate Eq.~\eqref{eq:validity} says that the approximation is better than $q=10\%$.
 }
 \label{fig:illustrate-delta-abmin}
\end{figure}

\begin{figure}
 \centering
 \includegraphics[width=\hsize]{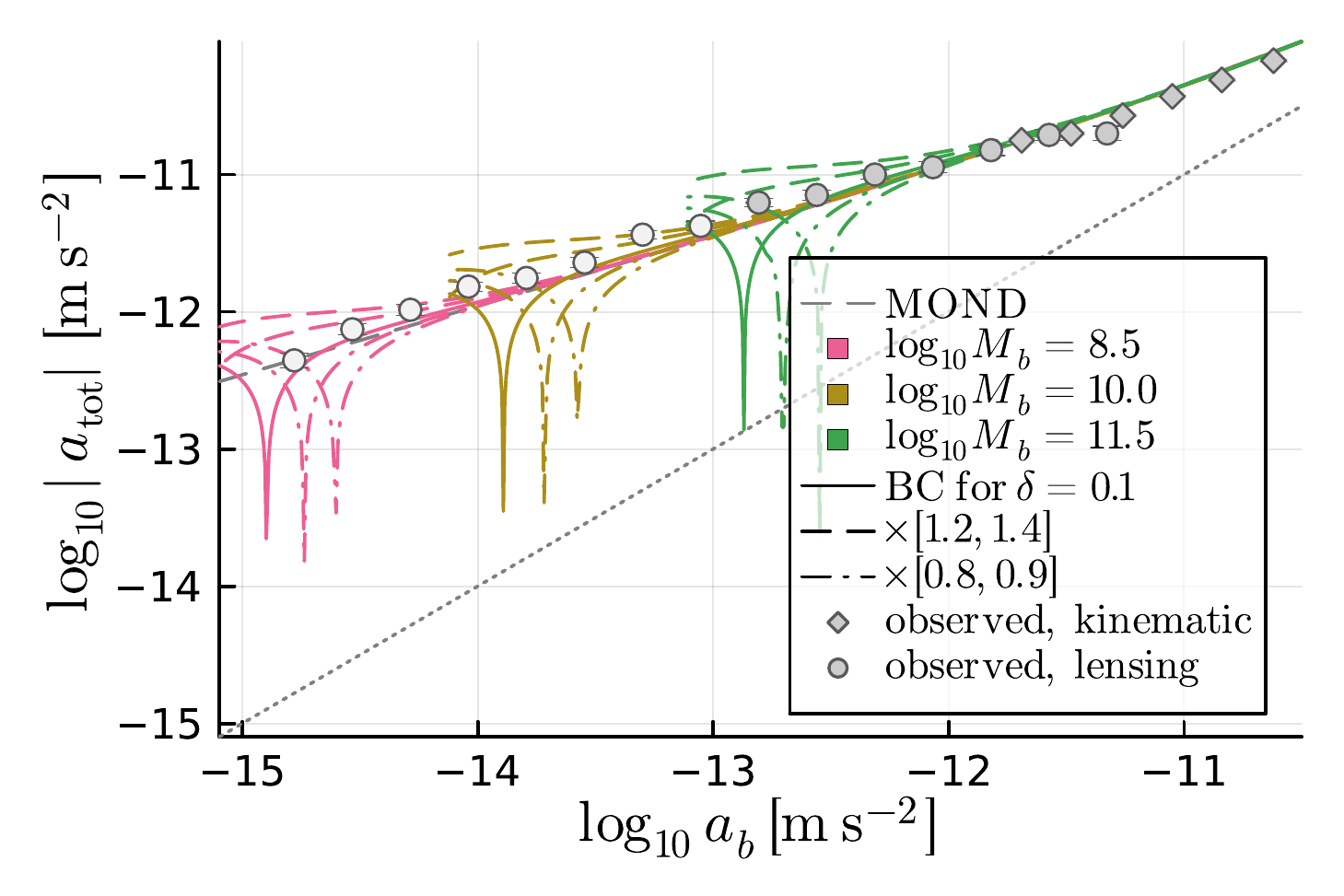}
 \caption{
   RAR for different baryonic masses and boundary conditions for numerical solutions with $f_G/m^2 = 0.99\,\mathrm{Mpc}^2$.
   The solid lines are for boundary conditions where $a_{\tilde{\varphi}}$ stays within a fraction $\delta = 0.1$ from MOND up to the largest possible radius for a given baryonic mass $M_b$.
   Dash-dotted and dashed lines correspond to these optimal boundary conditions multiplied by factors $[0.8, 0.9]$ and $[1.2, 1.4]$, respectively.
   The dips all go to $-\infty$ since they correspond to $M_{\mathrm{eff}} = 0$ (i.e., $a_{\mathrm{tot}} = 0$), but this is not resolved numerically.
   The y-axis shows the modulus of $a_{\mathrm{tot}}$.
   So after the dip, the direction of the accelerations is flipped.
   All solutions are cut off where the condensate density of the largest boundary-condition solution for a given $M_b$ first drops to zero.
   The observed weak-lensing RAR does not include the hot gas estimate of \citet{Brouwer2021}.
   We correct the observed weak-lensing RAR to be consistent with the $M/L_*$ scale of the observed kinematic RAR (McGaugh 2022, priv. comm.).
   Data points below $a_b = 10^{-13}\,\mathrm{m/s}^2$ have a lighter color since the isolation criterion is less reliable there \citep{Brouwer2021}.
 }
 \label{fig:illustrate-RAR}
\end{figure}

This maximum radius $r_{\mathrm{max}}$ scales as $M_b^{1/6}$. This means that more massive galaxies can stay close to MOND up to larger radii than less massive galaxies.
In MOND, however, accelerations $a_b = G_N M_b/r^2$ are more relevant than radii $r$.
For example, the Radial Acceleration Relation \cite[RAR,][]{Lelli2017b} relates the total acceleration $a_{\mathrm{tot}}$ and the Newtonian baryonic acceleration $a_b$.

The maximum radius $r_{\mathrm{max}}$ corresponds to a minimum acceleration $a_{b,\mathrm{min}} = G_N M_b/r_{\mathrm{max}}^2$ which scales as $M_b^{2/3}$.
Thus, in acceleration space, less massive galaxies can stay close to MOND for longer than more massive galaxies.
We show $a_{b,\mathrm{min}}$ in Fig.~\ref{fig:illustrate-delta-abmin} and illustrate the RAR for various baryonic masses in Fig.~\ref{fig:illustrate-RAR}.

The scale of $r_{\mathrm{max}}$ is set by the combination $m^2/f_G$.
\citet{Skordis2020} require $m^2/f_G \lesssim 1\,\mathrm{Mpc}^{-2}$.
As we show below, this does not guarantee MOND-like behavior for weak lensing which probes radii up to $\sim1\,\mathrm{Mpc}$.
One might therefore want to choose an even smaller $m^2/f_G$. But this is not easily possible.
Indeed,  galaxy clusters require more acceleration than \textsc{MOND} predicts \citep{Aguirre2001, Sanders2003,Eckert2022}.
To naturally explain this in the AeST model, $m^2/f_G$ cannot be much smaller than $1\,\mathrm{Mpc}^{-2}$ and we assumed
\begin{align}
 m^2/f_G \sim (1\,\mathrm{Mpc})^{-2} \,.
\end{align}
We obtained this estimate by requiring that the minimum acceleration $a_{b,\mathrm{min}}$ in AeST matches where observed clusters deviate from MOND.
For example, if we require that the total acceleration $a_{\mathrm{tot}}$ in a cluster with $M_b = 10^{14}\,M_\odot$ deviates by at least $\delta = 10\%$ from MOND at $a_{b,\mathrm{min}} = 10^{-10.5}\,\mathrm{m/s}^2$, we find $m^2/f_G > 2.5\,\mathrm{Mpc}^{-2}$ (see also Table~\ref{table:m2fGbounds}).\footnote{
  Here, we did not use the analytical estimate Eq.~\eqref{eq:rmax} for $r_{\mathrm{max}}$ when calculating $a_{b,\mathrm{min}} = G_N M_b/r_{\mathrm{max}}^2$.
  The reason is that Eq.~\eqref{eq:rmax} is derived using approximations that are better for galaxies than for clusters.
  Instead, we calculated $r_{\mathrm{max}}$ from numerical solutions of the equations of motion.
  Using Eq.~\eqref{eq:rmax} would give $m^2/f_G > 7.9\,\mathrm{Mpc}^{-2}$.
}

\begin{figure}
 \centering
 \includegraphics[width=\hsize]{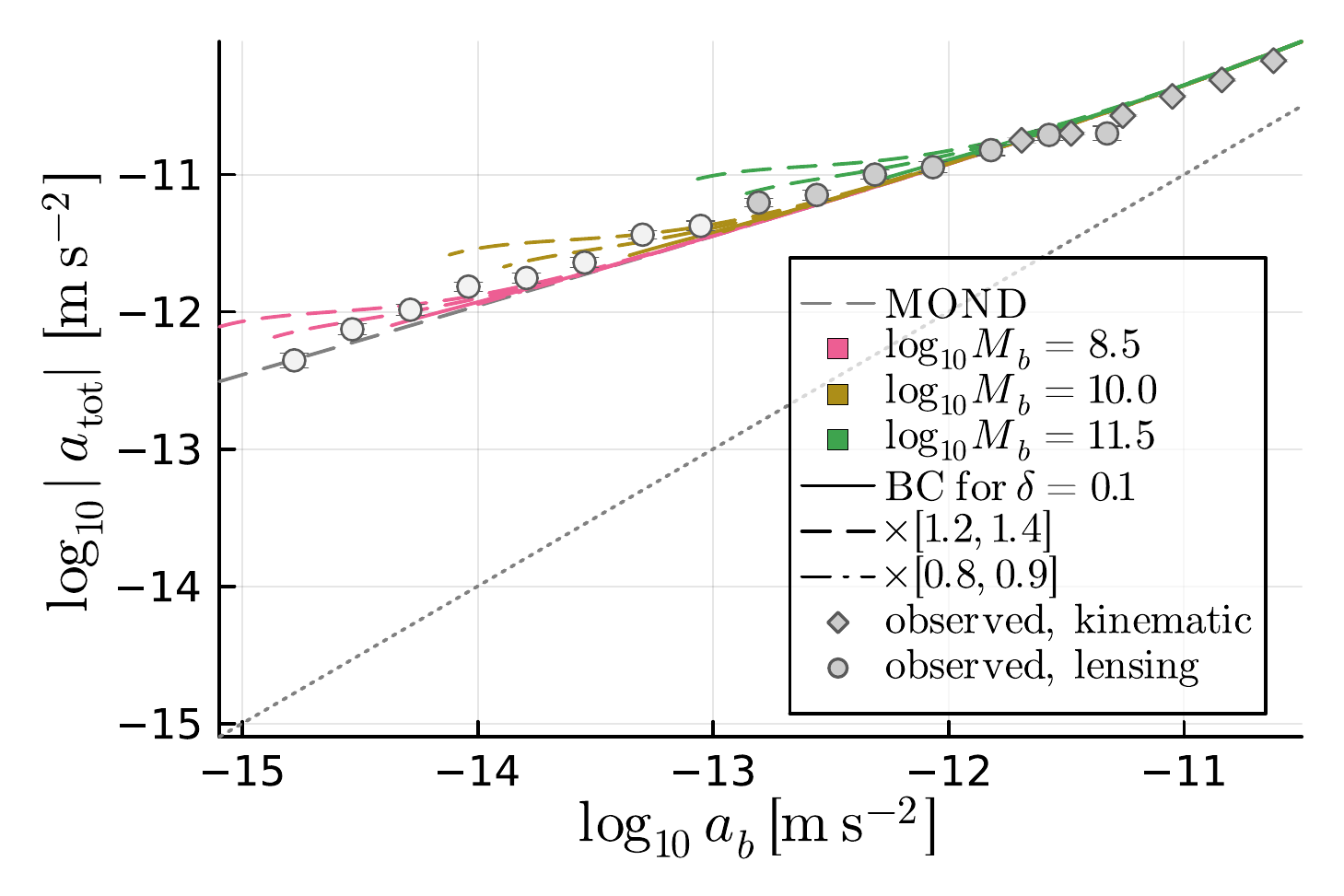}
 \caption{
   Same as Fig.~\ref{fig:illustrate-RAR} but with all solutions truncated where the condensate density first drops to zero, that is, truncated where the solutions become potentially unstable.
 }
 \label{fig:illustrate-RAR-posdensity}
\end{figure}

Beyond the radius where the AeST model deviates from MOND, the gravitational force eventually becomes oscillatory \citep{Skordis2020}, see also Appendix~\ref{sec:oscillations}, as is typical for condensate models \citep{Arkani-Hamed2007}.
The reason is that $M_{\mathrm{eff}}(r)$ and the condensate density $\rho_c$ oscillate.
However, condensates with negative energy-density are unstable (or at least we would expect a good reason for why they are not unstable).
We therefore expect the AeST model to be unstable in this oscillatory regime. We would then no longer have a macroscopically coherent condensate and the quasi-static action Eq.~\eqref{eq:action} is no longer good to use.

In superfluid dark matter models, for example, we assume that when the energy-density begins to oscillate that we have to continue the condensate density by a standard (nonsuperfluid) phase \citep{Berezhiani2015, Berezhiani2018}.
Something similar might happen in the AeST model.

However, so far, a stability analysis for the AeST model in a galactic background has not been done, so maybe the oscillatory regime turns out to be stable after all.
Below, we therefore keep in mind both possibilities and discuss where our results depend on whether or not negative condensate densities are stable.
For example,  Fig.~\ref{fig:illustrate-RAR-posdensity} shows the solutions from Fig.~\ref{fig:illustrate-RAR} but truncated where the condensate density first drops to zero.

\section{Weak lensing}

The AeST model usually reproduces MOND at the radii probed by rotation curves because the condensate density $M_c$ is negligible there.
This means that probing the effects of the condensate requires a different approach.
The option we pursued here is to use the recent weak-lensing analysis of \citet{Brouwer2021} who find that accelerations are MOND-like down to at least $a_b \sim 10^{-13}\,\mathrm{m}/\mathrm{s}^2$.

In the AeST model, matter is coupled to the fields $\varphi$ and $\hat{\Phi}$ through the metric $g_{\mu \nu}$.
In the weak-field limit, this metric has the same form as in GR, just with the Newtonian potential replaced by $\Phi = \hat{\Phi} + \varphi$.
We can therefore use the standard formalism for weak lensing just by taking into account $\Phi = \hat{\Phi} + \varphi$.

Most galaxies in the weak-lensing sample from \citet{Brouwer2021} have baryonic masses between $10^{10}\,M_\odot$ and $10^{11}\,M_\odot$.
For the AeST model with $f_G/m^2 = 0.99\,\mathrm{Mpc}^2$, Fig.~\ref{fig:illustrate-delta-abmin} shows that MOND-like behavior up to a few $10\%$ is possible down to $a_b \sim 10^{-13}\,\mathrm{m}/\mathrm{s}^2$ for those galaxies in the sample with baryonic masses  $\sim 10^{10}\,M_\odot$.

Such $\mathcal{O}(10\%)$ deviations may be sufficient to match observations.
But it is important to keep in mind that in this regime where deviations from MOND start to become important, the details depend on the precise baryonic masses and boundary conditions of the galaxy sample as well as the precise value of the model parameter $m^2/f_G$.
For example, Fig.~\ref{fig:illustrate-delta-abmin} shows solutions for boundary conditions that are optimal for reproducing MOND.
But there is no reason why galaxy formation should result in such optimal boundary conditions. One would therefore expect deviations of AeST from MOND to generally be larger than in the optimal case we depict.

\begin{table}
\caption{Rough bounds on $m^2/f_G$.}
\label{table:m2fGbounds}
\centering
\begin{tabular}{c c c c}
\hline\hline
\\[-0.9em] %
Bound on $m^2/f_G$ & Description \\
\hline
\\[-0.9em] %
$\lesssim 1 \,\mathrm{Mpc}^{-2}$ & Galaxies, weak lensing ($a_b \geq 10^{-13}\,\mathrm{m/s}^2$)\\
$\lesssim 0.001 \,\mathrm{Mpc}^{-2}$  & Galaxies, weak lensing ($a_b \geq 10^{-15}\,\mathrm{m/s}^2$)\\
$\gtrsim 1\,\mathrm{Mpc}^{-2}$ & Galaxy clusters ($a_b \sim 10^{-10.5}\,\mathrm{m/s}^2$)\\
\hline
\end{tabular}
\tablefoot{
  For the bounds from galaxies, we require that galaxies with $M_b = 10^{11}\,M_\odot$ can reproduce MOND up to a fraction $\delta = 10\%$ for the accelerations $a_b$ probed by weak lensing.
  That is, we require that the minimum acceleration $a_{b,\mathrm{min}}$ is sufficiently small for these galaxies.
  For the bounds for galaxy clusters, we require that a cluster with $M_b = 10^{14}\,M_\odot$ cannot reproduce MOND to better than $\delta = 10\%$ at accelerations $a_b$ a bit below the MOND acceleration scale $a_0$.
  That is, we require that $a_{b,\mathrm{min}}$ is sufficiently large for clusters.
  The bounds on $m^2/f_G$ scale roughly as $\delta \, a_{b,\mathrm{min}}^{3/2} /M_b$ for small $\delta$.
}
\end{table}

We can derive a rough upper bound on the model parameter $m^2/f_G$ by requiring that AeST can reproduce MOND in the regime probed by weak lensing.
For this, we used the minimum acceleration $a_{b,\mathrm{min}} = G_N M_b/r_{\mathrm{max}}^2$ with $r_{\mathrm{max}}$ from Eq.~\eqref{eq:rmax}.
Many of the galaxies in the sample used by \citet{Brouwer2021} have baryonic masses close to $10^{11}\,M_\odot$.
If we require that such galaxies can reproduce MOND down to $a_{b,\mathrm{min}} = 10^{-15}\,\mathrm{m/s}^2$ and up to a fraction $\delta = 10\%$, we find $m^2/f_G < 0.001\,\mathrm{Mpc}^{-2}$.
There is a lot of uncertainty in this upper bound.
For example, $m^2/f_G = 0.001\,\mathrm{Mpc}^{-2}$ is small enough that the ghost condensate density in galaxies as given by Eq.~\eqref{eq:rhoc} is typically smaller than the cosmological background density.
So, at least in principle, there could be corrections from the fact that we should expand around a cosmological background, not around empty Minkowski space (which is how Eq.~\eqref{eq:eom} was derived).\footnote{
  But we note that there is no reason to expect that such corrections would help to explain why weak-lensing observations follow the MOND prediction down to very small accelerations $a_b$.
}
Also, if we disregard the data below $a_b = 10^{-13}\,\mathrm{m/s}^2$ because the isolation criterion used in the weak-lensing analysis is less reliable there \citep{Brouwer2021}, we obtain the weaker bound $m^2/f_G < 1 \,\mathrm{Mpc}^{-2}$.
Still, a tension between the value of $m^2/f_G$ required by weak lensing and that required by galaxy clusters, $m^2/f_G \gtrsim 1\,\mathrm{Mpc}^{-2}$, seems likely (see Table~\ref{table:m2fGbounds}).

\begin{figure}
 \centering
 \includegraphics[width=\hsize]{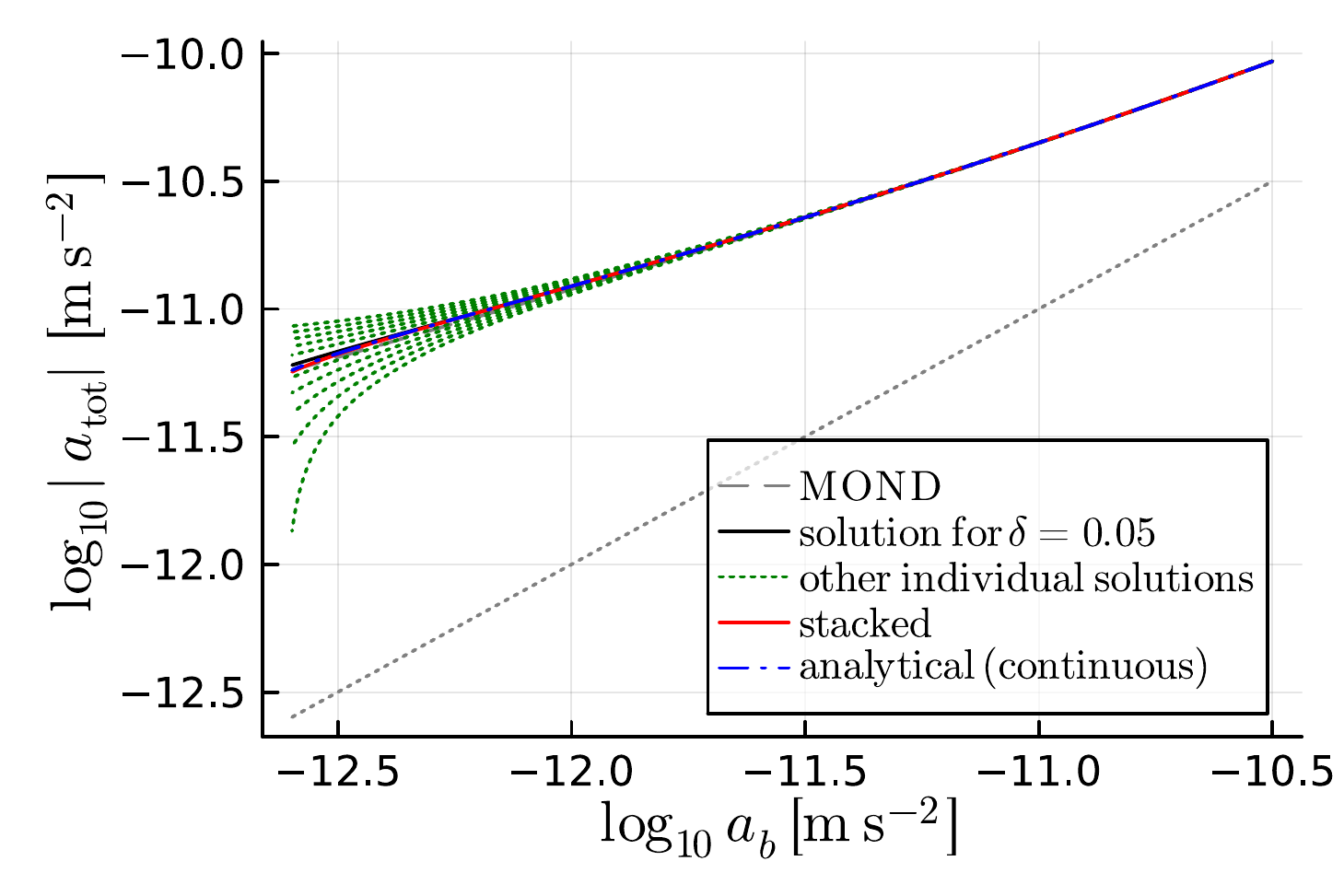}
 \caption{
   RAR for solutions with fixed baryonic mass $M_b = 10^{11}\,M_\odot$ for various boundary conditions (green dotted lines) and the corresponding stacked RAR (solid red line).
   This is for $f_G/m^2 = 0.99\,\mathrm{Mpc}^2$.
   The boundary conditions of the individual solutions are in the range $0.5 - 1.5$ times the optimal boundary condition for $\delta = 0.05$ with a step size of $0.1$ times the optimal one.
   Solutions are cut off where the first of the individual solutions reaches $a_{\mathrm{tot}} = 0$.
   We also show the analytically stacked RAR according to formula Eq.~\eqref{eq:atotstackedanalytical}.
 }
 \label{fig:illustrate-stacking-stopMeff}
\end{figure}

\begin{figure}
 \centering
 \includegraphics[width=\hsize]{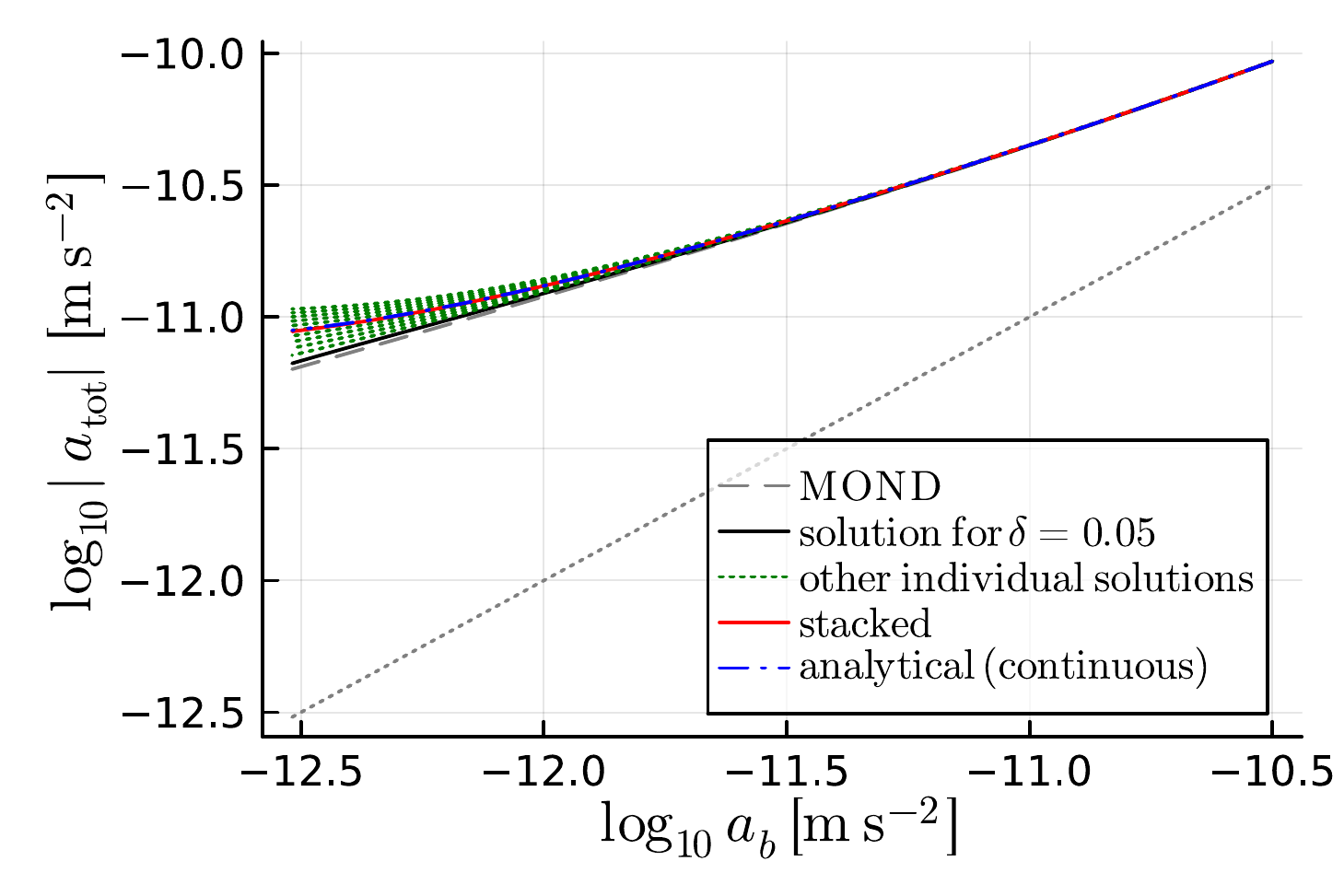}
 \caption{
   Same as Fig.~\ref{fig:illustrate-stacking-stopMeff} but with all individual stacked solutions having positive condensate density.
   The individual solutions are for boundary conditions in the range $1.0 - 2.0$ times the optimal boundary condition for $\delta = 0.05$ with a step size of $0.1$ times the optimal one.
   Solutions are cut off where the first of the individual solutions reaches $\rho_c = 0$.
 }
 \label{fig:illustrate-stacking-stoprhocond}
\end{figure}

Another aspect to take into account is that, in practice, the weak-lensing RAR is not known for individual galaxies.
It is known only in an averaged sense for a large sample of stacked galaxies.
It is possible that stacking gives a MOND-like RAR
  even if most galaxies individually do not.
As we show in Appendix~\ref{sec:appendix:stacking}, in our case, stacking simply means calculating a weighted average in acceleration space.
So, indeed, accelerations larger than MOND from some galaxies can cancel accelerations smaller than MOND from other galaxies.

To illustrate this, we considered a sample of galaxies with a fixed baryonic mass $M_b$ but various boundary conditions.
For simplicity, we weighed all galaxies equally when averaging.
One example is shown in Fig.~\ref{fig:illustrate-stacking-stopMeff}.
We see how a stacked RAR can be MOND-like even when most individual stacked galaxies are not.
Of course, one does not always get a MOND-like RAR from stacking.
This works only when accelerations below and above the MOND prediction cancel each other.
Whether or not it works for real lensing galaxies depends on which boundary conditions are picked by galaxy formation.

In addition, stacking galaxies with different boundary conditions should lead to increased uncertainties at small $a_b$ where different boundary conditions lead to significantly different accelerations.
In principle, such uncertainties might be visible in the error bars of the observed weak-lensing RAR.

However, we expect that in practice there is often no visible effect on the error bars.
To see this, we first note that the error bars shown in Fig.~\ref{fig:illustrate-RAR} and Fig.~\ref{fig:illustrate-RAR-posdensity} correspond to the uncertainty of the mean value of $a_{\mathrm{tot}}$, obtained by stacking a large number $N$ of galaxies.
They do not represent the scatter from galaxy-by-galaxy variation.
The galaxy-by-galaxy variation is larger than the uncertainty of the mean by a factor of about $\sqrt{N}$.
In AeST, different boundary conditions induce a form of galaxy-by-galaxy variation.
Thus, the scatter from solutions with different boundary conditions should be compared to the larger galaxy-by-galaxy variation and not to the uncertainty of the mean.
Put differently, the scatter induced by different boundary conditions should be scaled by a factor $1/\sqrt{N}$ before comparing to the error bars shown in Fig.~\ref{fig:illustrate-RAR} and Fig.~\ref{fig:illustrate-RAR-posdensity}.
Here, $N$ is on the order of $10^5$ \citep{Brouwer2021}.
Thus, one would expect a visible effect on the error bars only in extreme cases.

\subsection{Negative condensate densities}

Whether or not stacked galaxies can follow a MOND RAR down to smaller accelerations than individual galaxies
  also depends on whether or not negative condensate densities are stable.
To see this, we note that accelerations are smaller than in MOND if and only if the effective mass $M_{\mathrm{eff}} = M_b + M_c$ is smaller than in MOND.
This is only possible for negative $M_c$ which requires negative condensate densities.
Therefore, if negative densities are unstable, cancelling accelerations that are larger against those that are smaller than in MOND does not work.
This is simply because all accelerations are larger than in MOND if we do not allow for negative densities.
This is illustrated in Fig.~\ref{fig:illustrate-stacking-stoprhocond}.

\begin{figure}
 \centering
 \includegraphics[width=\hsize]{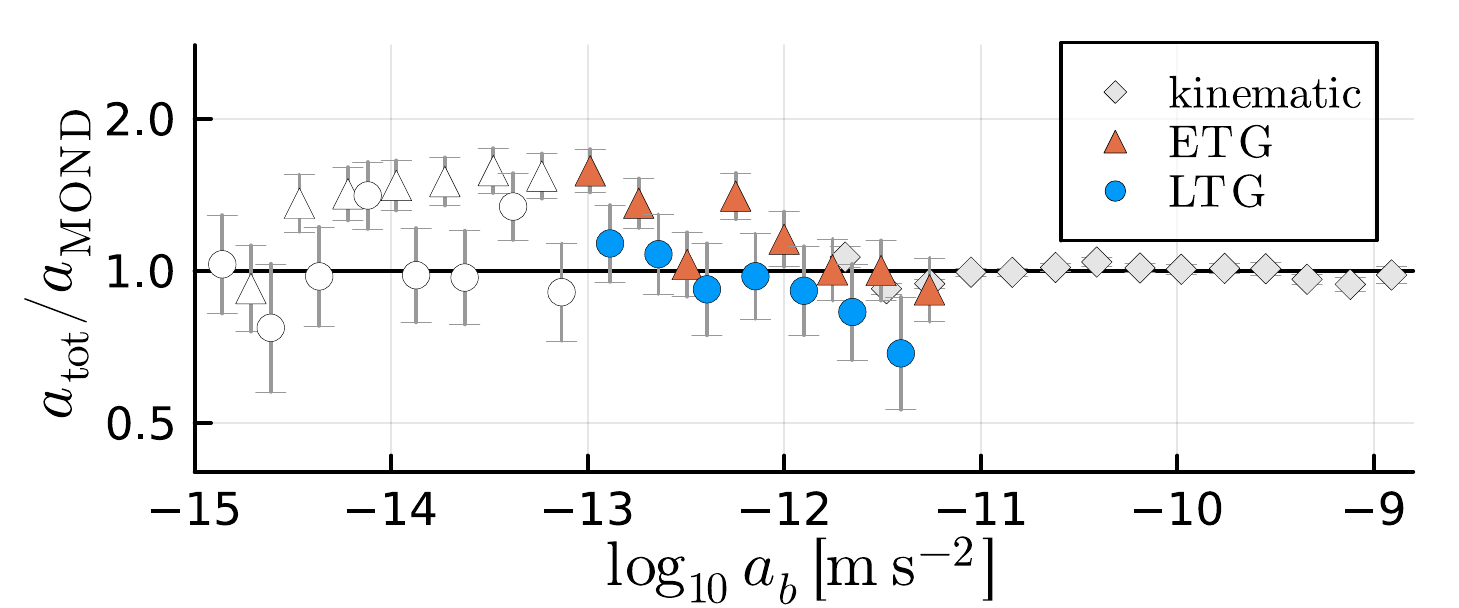}
 \caption{
   Observed weak-lensing RAR for ETGs and LTGs from \citet{Brouwer2021} with the stellar $M/L_*$ corrected to use the same stellar population model as the observed kinematic RAR (McGaugh 2022, priv. comm.) relative to the MOND prediction.
   This does not include the hot gas estimate from \citet{Brouwer2021}.
   Here, we take $a_{\mathrm{MOND}} = a_b \nu_e(a_b/a_0)$ with $\nu_e(y) = (1 + e^{-\sqrt{y}})^{-1}$ \citep{Lelli2017b}.
   Data points below $a_b = 10^{-13}\,\mathrm{m/s}^2$ are shown in white since the isolation criterion is less reliable there \citep{Brouwer2021}.
 }
 \label{fig:lensing-RAR-ETG-vs-LTG}
\end{figure}

Thus, if negative densities are unstable, one might expect that the AeST model always gives larger accelerations than MOND.
Moreover, our estimate for $a_{b,\mathrm{min}}$ (see Fig.~\ref{fig:illustrate-delta-abmin}) suggests that these deviations should set in earlier for larger baryonic masses $M_b$.
And indeed, there are hints of such behavior in the observed weak-lensing data, see Fig.~\ref{fig:lensing-RAR-ETG-vs-LTG} which shows the weak-lensing data separately for early-type galaxies (ETGs) and late-type galaxies (LTGs).
We see that the weak-lensing RAR for LTGs follows the MOND prediction even for $a_b < 10^{-13}\,\mathrm{m/s}^2$ while ETGs tend toward larger accelerations than MOND.
In general, ETGs have larger baryonic masses than LTGs.
So this seems to fit with the AeST model expectations if negative densities are unstable.

However, this $M_b$-dependence is not a plausible explanation for the difference between the observed weak-lensing RARs for ETGs and LTGs. This is for three reasons.

First, the ETGs and LTGs do not sufficiently differ in baryonic mass.
What would be required is a difference in $M_b$ of more than a factor $10^{3/2}$.
To see this, we note that LTGs follow the MOND prediction for at least one more order of magnitude in $a_b$ compared to ETGs.
This translates into a factor $>10^{3/2}$ in terms of baryonic mass according to our estimate $a_{b,\mathrm{min}} \propto M_b^{2/3}$.
In contrast, \citet{Brouwer2021} selected LTGs and ETGs to have the same stellar mass distribution.
With the stellar $M/L$ scale corrected to be consistent with that of the observed kinematic RAR, ETG stellar masses would still be larger by a factor $1.4$ (McGaugh 2022, priv. comm.).
But this is not sufficient here.

Of course, the total baryonic mass should take into account gas, but this is unlikely to account for the required factor of $10^{3/2}$ or more, at least with the simple cold gas mass estimate used in \citet{Brouwer2021}.
\citet{Brouwer2021} also consider a scenario where ETGs have significantly more hot gas than LTGs.
But even in that scenario, the baryonic masses of ETGs differ from those of LTGs by only a factor of two or a bit more.
In addition, adopting this scenario means adopting a different observed lensing RAR.
Indeed, as discussed in \citet{Brouwer2021}, in this scenario there might not even be a discrepancy between ETGs and LTGs left to explain.

Second, the isolation criterion needed to obtain the weak-lensing RAR might fail for ETGs sooner than it does for LTGs.
One might naturally expect this to be the case since ETGs are known to be more clustered than LTGs \citep{Dressler1980}.
At what point this comes into play here we cannot judge, but mention it as a logical possibility.

Third and finally, even if negative densities are indeed unstable, the AeST model does not necessarily predict larger accelerations than MOND.
Indeed, it makes no physical sense to stop looking when the condensate becomes unstable.
In a real galaxy, something else must follow after the condensate phase.
For example, the macroscopically coherent ghost condensate might be replaced by something closer to a $\Lambda$CDM-like collisionless fluid which the AeST model postulates on cosmological scales \citep{Skordis2020}.
In principle, whatever replaces the ghost condensate might lead to smaller accelerations than MOND.

That said, the prediction of larger-than-MOND accelerations remains valid if whatever replaces the ghost condensate has as its only effect to replace the ghost condensate density by some other positive density, $\rho_c \to \rho_{\mathrm{replace}}$.
This is because then one still has solutions of the same form as before, just with a different effective mass $M_{\mathrm{eff}} \quad \to \quad M_b + M_{\mathrm{replace}} > M_b$.

In order to get smaller-than-MOND accelerations, the general structure of the solutions must be modified.
That is, the left-hand sides of the equations of motions must be modified,
$ \Delta \hat{\Phi} = \dots \,, \vec{\nabla} \left(\tilde{\mu}(|\vec{\nabla} \varphi|/a_0) \vec{\nabla} \varphi\right) = \dots \to \, ?$.
It is not implausible that this indeed happens since the field $\varphi$ plays a role for both the ghost condensate (for example, it carries the chemical potential $\dot{\varphi} = \mu$) as well as the gravitational force.
So outside the condensate phase both could be modified.

\subsection{External field effect}

Another concern at the small accelerations probed by weak lensing is the external field effect (EFE) of MOND \citep{Milgrom1983a, Famaey2012}.
The EFE is a consequence of the specific nonlinear form $\vec{\nabla} (|\vec{\nabla} \varphi| \vec{\nabla} \varphi) \propto \rho_b$ of the gravitational field equations in the small-acceleration limit.
The crucial nonlinearity is the same in the AeST model, so there is probably a similar effect there,
  at least within the ghost condensate.\footnote{
    The EFE is only relevant in situations that are not spherically symmetric.
    In these cases, as we discuss in Appendix~\ref{sec:appendix:Azeroassumption}, the vector field $\vec{A}$ of the AeST model cannot be set to zero and the equations of motion are not given by Eq.~\eqref{eq:eom}.
    Still, there is the same type of nonlinearity and it is plausible that an effect similar to the MOND EFE exists.
  }

The EFE generally reduces the observationally inferred $a_{\mathrm{tot}}$,
 while a condensate mass $M_c > 0$ in the AeST model enhances this acceleration.
In principle, these two effects could cancel each other to give a MOND-like acceleration even at very large galactocentric distances, but there is no reason to expect such a cancellation to generally happen.

And in any case, if negative densities are unstable and the condensate is replaced by something else at large radii, then any potential EFE depends on the details of what replaces the ghost condensate.
There could be a modified nonlinear effect that still allows neighboring galaxies to affect each other in a way that violates the Strong Equivalence Principle (SEP) like the MOND EFE does.
Or there could be no such effect, possibly restoring the SEP at large scales.

A restored SEP would fit with the fact that the observed weak-lensing RAR from \citet{Brouwer2021} shows no signs of an EFE.
But here one must be careful.
The EFE pertains to nonisolated galaxies, while the analysis of \citet{Brouwer2021} requires isolated galaxies.
Thus, any EFE effects may be masked by violations of this assumption.
In addition, the environment-dependence of the EFE is quite complicated \citep{Llinares2008, Chae2021}.
So it is not even clear for MOND whether or not a significant EFE is expected here.
Still, the EFE is something to keep in mind as observations and theoretical predictions are improved.

\section{Conclusion}

We have explored whether or not the AeST model can explain the observed MOND-like weak-lensing RAR which probes unprecedentedly small accelerations.
We find that deviations from MOND start to set in already in the range of the new measurements, creating a tension with data.
The model parameter $m^2/f_G$ can be adjusted to avoid this tension, but that likely creates a tension with observations of galaxy clusters instead.

It seems that keeping the model in agreement with data would require specific values of boundary conditions for a large variety of galaxies.
While this is possible, we do not know of any mechanism that would result in these particular boundary conditions.
Thus, while we cannot rule out the model, it does seem that weak-lensing observations pose a challenge for AeST.

\begin{acknowledgements}
We thank Margot Brouwer, Andrej Dvornik, and Kyle Oman for helpful correspondence.
This work was supported by the DFG (German Research Foundation) under grant number HO 2601/8-1 together with the joint NSF grant PHY-1911909.
This work was supported by the DFG (German Research Foundation) – 514562826.
\end{acknowledgements}

\bibliographystyle{aa} %
\bibliography{AeST-lensing.bib} %

\begin{appendix}

\section{Action and equations of motion}
\label{sec:appendix:actionandeom}

For galaxies, the quasi-static weak-field limit of the AeST model is relevant.
The action in this limit is \citep{Skordis2020}
\begin{multline}
 \label{eq:action}
 S = - \int d^4x \left\{\frac{1}{8 \pi \hat{G}} \left[ (\vec{\nabla} \Phi)^2 - 2 \vec{\nabla} \Phi \vec{\nabla} \varphi + (\vec{\nabla} \varphi)^2 \right.\right.\\
\left.\left. - m^2 \left(\frac{\dot{\varphi}}{Q_0} - \Phi\right)^2 + \mathcal{J}\left((\vec{\nabla} \varphi)^2\right) \right] + \Phi \rho_b \right\} \,,
\end{multline}
where $\hat{G}$ is a constant and the function $\mathcal{J}$ determines the MOND interpolation function.
We discuss potential higher-order corrections to the $m^2$ term which produces the ghost condensate density in Appendix~\ref{sec:appendix:higherorder}.

The AeST model also contains a unit vector field $A^\mu$.
The form Eq.~\eqref{eq:action} of the action assumes $\vec{A} = 0$ \citep{Skordis2020}.
In Appendix~\ref{sec:appendix:Azeroassumption} we explain why this assumption is correct in spherical symmetry -- which is what we are mainly interested in here -- but not in general.

The quasi-static weak-field limit equations of motion derived from the action Eq.~\eqref{eq:action} are
\begin{subequations}
\begin{align}
\Delta \hat{\Phi} &= 4 \pi \hat{G} \rho_b + m^2(\mu/Q_0 -\hat{\Phi} - \varphi) \,, \\
\vec{\nabla} \left( \tilde{\mu}\left(\frac{|\vec{\nabla} \varphi|}{a_0}\right) \vec{\nabla} \varphi \right) &= 4 \pi \hat{G} \rho_b + m^2(\mu/Q_0 -\hat{\Phi} - \varphi) \,,
\end{align}
\end{subequations}
where, following \citet{Skordis2020}, we have introduced $\hat{\Phi}$ through $\Phi \equiv \hat{\Phi} + \varphi$ and $\tilde{\mu}(|\vec{\nabla} \varphi|/a_0) = \mathcal{J}'((\vec{\nabla} \varphi)^2)$.
In order to reproduce MOND-like behavior for small accelerations and Newton-like behavior for large accelerations, the function $\tilde{\mu}(|\vec{\nabla} \varphi|/a_0)$ must be proportional to $|\vec{\nabla} \varphi|$ for small arguments and it must be a constant for large arguments.
One can parametrize these limits by a parameter $\lambda_s$ \citep{Skordis2020},
\begin{align}
 \left.\tilde{\mu}\left(\frac{|\vec{\nabla} \varphi|}{a_0}\right)\right|_{|\vec{\nabla} \varphi| \to 0} = \frac{\lambda_s}{1 + \lambda_s} \frac{|\vec{\nabla} \varphi|}{a_0} \,, \quad
 \left.\tilde{\mu}\left(\frac{|\vec{\nabla} \varphi|}{a_0}\right)\right|_{|\vec{\nabla} \varphi| \to \infty} = \lambda_s \,.
\end{align}
This ensures both a standard Newton regime at large accelerations and a standard MOND regime at small accelerations with Newtonian gravitational constant
\begin{align}
 G_N \equiv \hat{G} \, \frac{1 + \lambda_s}{\lambda_s} \equiv \hat{G} \cdot f_G^{-1} \,.
\end{align}

How exactly the function $\tilde{\mu}$ interpolates between these two limits is not specified by the AeST model.
Various choices are possible.
A choice of $\tilde{\mu}$ corresponds to a choice of the so-called interpolation function in MOND \citep{Famaey2012}.
Up to the condensate density $\rho_c$, these equations are typical of multifield MOND models \citep{Famaey2012}.

In the small-acceleration limit $|\vec{\nabla} \varphi| \ll a_0$ and without any baryonic density, $\rho_b = 0$, these equations approximately become $\vec{\nabla} (|\vec{\nabla} \varphi| \vec{\nabla} \varphi) = c_1 + c_2 \varphi$ with constants $c_1$ and $c_2$.
This is a special case of the deep-MOND polytropes studied in \citet{Milgrom2021b}.
The results of \citet{Milgrom2021b} are not directly useful here, however, because the baryonic density $\rho_b$ plays a very important role for the phenomenology of the AeST model as we see below.

\section{The quasi-static limit more generally}
\label{sec:appendix:Azeroassumption}

Had we not set $\vec{A}$ to zero by hand, the action in the quasi-static limit Eq.~\eqref{eq:action} would read,
\begin{multline}
 \label{eq:actionwithA}
 S = - \int d^4x \left\{\frac{1}{8 \pi \hat{G}} \left[ (\vec{\nabla} \Phi)^2 - 2 \vec{\nabla} \Phi \, (\vec{\nabla} \varphi + Q_0 \vec{A}) + (\vec{\nabla} \varphi + Q_0 \vec{A})^2 \right.\right.\\
\left.\left. - m^2 \left(\frac{\dot{\varphi}}{Q_0} - \Phi\right)^2 + \mathcal{J}\left((\vec{\nabla} \varphi + Q_0 \vec{A})^2\right) +
\frac{2K_B}{2-K_B} \vec{\nabla}_{[i} \vec{A}_{j]}  \vec{\nabla}^{[i} \vec{A}^{j]}  \right] + \Phi \rho_b \right\} \,.
\end{multline}
We can decompose $\vec{A}$ into a divergence-less and a curl-less part, $\vec{A} \equiv \vec{\nabla} \times \vec{A}_c + \vec{\nabla} A_d$.
For time-independent fields -- that is, time-independent up to the chemical potential $\dot{\varphi} = \mu = \mathrm{const}.$ -- the results of \citet{Skordis2021} show that we can set $A_d = 0$ by a gauge transformation.

In spherical symmetry, the curl term $\vec{\nabla} \times \vec{A}_c$ vanishes and the $\vec{A}$ equation of motion and the $\varphi$ equation of motion are equivalent.
Thus, setting $\vec{A}$ to zero and using Eq.~\eqref{eq:action} is justified.
In general, however, setting $\vec{A}$ to zero is inconsistent.
This can be seen from the $\vec{A}$ equation of motion,
\begin{multline}
\vec{\nabla} \Phi + \frac1{2 Q_0} \frac{2K_B}{2-K_B} \left( \Delta \vec{A} - \vec{\nabla} (\vec{\nabla} \cdot \vec{A}) \right)=\\
(\vec{\nabla} \varphi + Q_0 \vec{A}) \left(1 + \mathcal{J}'\left((\vec{\nabla} \varphi + Q_0 \vec{A})^2\right)\right) \,.
\end{multline}
If $\vec{A}$ were zero, we could infer
\begin{align}
 \label{eq:curlwithoutA}
 \vec{\nabla} |\vec{\nabla} \Phi| \times \vec{\nabla} \Phi = 0 \,,
\end{align}
by algebraically solving for $\vec{\nabla} \varphi$ and then taking the curl.
But Eq.~\eqref{eq:curlwithoutA} holds only in very special situations, see for example \citet{Brada1995}.
Thus, except in a few special cases, we must not set $\vec{A} = 0$ in the quasi-static limit.

Of course, even when Eq.~\eqref{eq:curlwithoutA} does not hold, setting $\vec{A}$ to zero might still be a reasonable approximation akin to how it is often reasonable to neglect a curl term in standard models of MOND \citep{Famaey2012}.
Investigating this is left for future work.

\section{General structure of the solutions}
\label{sec:appendix:solutions}

We now assume spherical symmetry.
Then, solutions have the same form as in standard multifield MOND models except that the baryonic mass $M_b$ is replaced by the effective mass $M_{\mathrm{eff}}$ which includes the condensate mass $M_c$ in addition to $M_b$.
The equations of motion are then solved by
\begin{subequations}
\begin{align}
 a_{\hat{\Phi}} \equiv \hat{\Phi}'(r) &= f_G \, a_N(r) \,, \\
 a_\varphi \equiv \varphi'(r) &= a_N(r) \tilde{\nu}\left(\frac{|a_N(r)|}{a_0}\right) \,,
\end{align}
\end{subequations}
where $a_N = G_N M_{\mathrm{eff}}(r)/r^2$ and the function $\tilde{\nu}$ is determined by $\tilde{\mu}$.
See, for example, \citet{Famaey2012} for how these two are related.
The total acceleration in the negative radial direction felt by matter is then
\begin{align}
\begin{split}
a_{\mathrm{tot}}(r)
 &= \hat{\Phi}'(r) + \varphi'(r) = a_N(r) \cdot \left(f_G + \tilde{\nu}\left(\frac{|a_N(r)|}{a_0}\right) \right)\\
 &\equiv a_N(r) \cdot \nu\left(\frac{|a_N(r)|}{a_0}\right) \,,
\end{split}
\end{align}
where $\nu$ is a MOND interpolation function \citep{Famaey2012}.

Below, we are mainly interested in the deep-MOND regime $a_N \ll a_0$.
In this regime, the total acceleration does not depend much on the choice of interpolation function.
Thus, we chose an interpolation function that is easy to handle,
\begin{align}
 \nu(y) \equiv 1 + 1/\sqrt{y} \,.
\end{align}
This interpolation function is not, in general, suited to fit galaxies \citep{Famaey2012}.
But it is sufficient in the small-acceleration limit in which we are interested here.
This implies for the accelerations due to the field $\hat{\Phi}$ and $\varphi$,
\begin{subequations}
\label{eq:dphi}
\begin{align}
 \hat{\Phi}'(r) &= \signeff \cdot f_G \, |a_N(r)| \,, \\
 \varphi'(r) &= \signeff \cdot \left( (1 - f_G) \, |a_N(r)| + \sqrt{a_0 |a_N(r)|} \right) \,,
\end{align}
\end{subequations}
and therefore for the total acceleration
\begin{align}
\begin{split}
 \label{eq:atotnunaive}
 \hat{\Phi}'(r) + \varphi'(r) &= \signeff \cdot \left( |a_N(r)| + \sqrt{a_0 |a_N(r)|}  \right) \\
 &= \signeff \cdot \left(\frac{G_N |M_{\mathrm{eff}}(r)|}{r^2} + \frac{\sqrt{a_0 G_N |M_{\mathrm{eff}}(r)|}}{r}\right)\,.
\end{split}
\end{align}
For later use, we define
\begin{align}
 \tilde{\Phi}'(r) \equiv \signeff \cdot |a_N(r)| \,, \quad \tilde{\varphi}'(r) \equiv \signeff \cdot \sqrt{a_0 |a_N(r)|} \,.
\end{align}
That is, $\tilde{\Phi}$ carries the Newton-like part of the total acceleration and $\tilde{\varphi}$ carries  the MOND-like part.
The total acceleration can be calculated from either $\tilde{\Phi} + \tilde{\varphi}$ or $\hat{\Phi} + \varphi$ since their sum is the same,
\begin{align}
 \hat{\Phi} + \varphi = \tilde{\Phi} + \tilde{\varphi} \,.
\end{align}

\section{Approximate analytical solution}
\label{sec:approximation}

The formal solutions Eq.~\eqref{eq:dphi} are not directly useful since the effective mass $M_{\mathrm{eff}}$ in $a_N$ depends on the value of the fields themselves through $\hat{\Phi} + \varphi$.
Equation~\eqref{eq:dphi} can, however, be used to recursively solve for $M_{\mathrm{eff}}$ starting at small radii where the baryonic mass $M_b$ dominates.
We first derive the recursion formula and then use it to obtain a first order approximation for $M_{\mathrm{eff}}$.
For simplicity, we assume a point-particle baryonic mass distribution, $\rho_b(\vec{x}) = M_b \delta(\vec{x})$.

In the following, we split the fields into a part due only to baryons, that is, $\hat{\Phi}_b$ and $\varphi_b$, and the rest, $\varphi_c$ and $\hat{\Phi}_c$ which includes the effects of the ghost condensate,
\begin{subequations}
\begin{align}
\hat{\Phi} \equiv \hat{\Phi}_b + \hat{\Phi}_c \,, \\
\varphi \equiv \varphi_b + \varphi_c \,,
\end{align}
\end{subequations}
where
\begin{subequations}
\begin{align}
 \hat{\Phi}_b &\equiv - f_G \, \frac{G_N M_b}{r} \,, \\
 \varphi_b &\equiv - (1 - f_G) \frac{G_N M_b}{r} +  \sqrt{G_N a_0 M_b} \ln(r/l) \,,
\end{align}
\end{subequations}
for some $l$.
This split depends on the additive constants chosen for $\hat{\Phi}_b$ and $\varphi_b$.
In particular, it depends on the choice of $l$ which parametrizes the additive constant in $\varphi_b$.

These additive constants can be shuffled around arbitrarily within the physical combination
\begin{align}
 \mu/Q_0 - \hat{\Phi}_b - \hat{\Phi}_c - \varphi_b - \varphi_c \,.
\end{align}
To solve the equations of motions, we need to impose a boundary condition for this combination.
In practice, we first fixed a value of $l$, that is, we fixed the additive constant in $\varphi_b$.
Then, it is equivalent to impose a boundary condition for $\mu/Q_0 - \hat{\Phi}_c - \varphi_c$.
Here, we chose to impose a value for this combination at $r=0$,
\begin{align}
 \mu/Q_0 - \hat{\Phi}_c(0) - \varphi_c(0) \,.
\end{align}

The effective mass $M_{\mathrm{eff}}$ is the sum of the baryonic mass $M_b$ and the integrated ghost condensate density $\rho_c$ which depends on the combination $\mu/Q_0 - \varphi - \hat{\Phi}$, see Eq.~\eqref{eq:rhoc}.
In turn, Eq.~\eqref{eq:atotnunaive} allows to calculate the derivative of this combination from $M_{\mathrm{eff}}$.
One can use this to derive a recursion formula for $M_{\mathrm{eff}}$ by equating two different expressions for $\mu/Q_0 -\varphi - \hat{\Phi}$.

In order to get $\mu/Q_0 - \varphi - \hat{\Phi}$ from the derivatives from Eq.~\eqref{eq:atotnunaive}, one must integrate once,
\begin{multline}
 \mu/Q_0 - \varphi(r) - \hat{\Phi}(r) = \mu/Q_0 - \varphi_b(r) - \hat{\Phi}_b(r) - \varphi_c(0) - \hat{\Phi}_c(0)\\- \int_0^r dr' ( \varphi'(r') + \hat{\Phi}'(r) - \varphi_b'(r') - \hat{\Phi}_b'(r'))\,.
\end{multline}
On the right-hand side, we can plug in Eq.~\eqref{eq:atotnunaive}, that is, $\varphi' + \hat{\Phi'} = \signeff (G_N |M_{\mathrm{eff}}|/r^2 + \sqrt{a_0 G_N |M_{\mathrm{eff}}|}/r)$, and the same expression but with $M_b$ instead of $M_{\mathrm{eff}}$ for $\varphi_b' + \hat{\Phi}_b'$.
The left-hand side is proportional to the condensate density $\rho_c$.
Thus, after multiplying by $r^2$ and integrating once more, the left-hand side is $M_{\mathrm{eff}}$.
We find,
\begin{multline}
 \label{eq:Meffrecursion}
 \frac{M_{\mathrm{eff}}(x)}{M_b} = 1 + \alpha \int_0^x dx' x'^2 \left\{
    p_l - \frac{1}{a_0 r_{\mathrm{MOND}}}(\varphi_b(x')  + \hat{\Phi}_b(x'))
    \right. \\ \left.
    - \int_0^{x'} dx''  \left[
      \frac{1}{x''^2} \left(\frac{M_{\mathrm{eff}}(x'')}{M_b}-1\right)
      \right. \right. \\ \left. \left.
      + \frac{1}{x''} \left(\signeff(x'')\sqrt{\frac{|M_{\mathrm{eff}}(x'')|}{M_b}}-1\right)
    \right]
   \right\} \,,
\end{multline}
where $r_{\mathrm{MOND}} = \sqrt{G_N M_b/a_0}$ is the MOND radius. We further defined
\begin{align}
 x \equiv \frac{r}{r_{\mathrm{MOND}}} \,, \quad \alpha \equiv \frac{m^2}{f_G}  r_{\mathrm{MOND}}^2\,,
\end{align}
and the parameter $p_l$ encodes the boundary condition,
\begin{align}
p_l \equiv \frac{1}{a_0 r_{\mathrm{MOND}}} \left(\frac{\mu}{Q_0} - \varphi_c(0) - \hat{\Phi}_c(0)\right) \,.
\end{align}
The subscript $l$ indicates that the split between $\varphi_b$, $\hat{\Phi}_b$ and $\varphi_c$, $\hat{\Phi}_c$ depends on $l$.

Equation~\eqref{eq:Meffrecursion} can be used as a recursion formula to iteratively solve for $M_{\mathrm{eff}}$.
The zeroth order approximation is to forget about the ghost condensate density, corresponding to $\alpha = 0$, and set
\begin{align}
 M_{\mathrm{eff},0} \equiv M_b \,.
\end{align}
The first order approximation is obtained from Eq.~\eqref{eq:Meffrecursion} by using the zeroth order approximation on the right-hand side (i.e., by setting $M_{\mathrm{eff}} = M_{\mathrm{eff},0} = M_b$ there),
\begin{align}
 \label{eq:Meffx}
\begin{split}
 M_{\mathrm{eff},1}(x) &\equiv M_b \left[ 1 + \alpha \int_0^x dx' x'^2 \left(
    p_l - \frac{1}{a_0 r_{\mathrm{MOND}}}(\varphi_b(x')  + \hat{\Phi}_b(x'))\right)\right] \\
    &= M_b \left[1 + \alpha x^2 \left(\frac12 + \frac19 x \left(3p  + 1 - 3 \ln(x)\right)\right) \right] \,,
\end{split}
\end{align}
where
\begin{align}
 p \equiv p_l - \ln\left(\frac{r_{\mathrm{MOND}}}{l}\right) = p_{l = r_{\mathrm{MOND}}} \,.
\end{align}
That is, $p$ is the boundary condition for $\mu/Q_0 - \varphi_c - \hat{\Phi}_c$ for the choice $l = r_{\mathrm{MOND}}$.

This first order estimate can also be obtained more directly by setting $\varphi = \varphi_b$ and $\hat{\Phi} = \hat{\Phi}_b$ in the ghost condensate density Eq.~\eqref{eq:rhoc}.
The advantage of the recursion formula Eq.~\eqref{eq:Meffrecursion} is that it is, at least conceptually, straightforward to improve on this approximation.
And it allows to analytically estimate when the first order approximation breaks down by going to the next higher order, see Appendix~\ref{sec:approx:validity}.

In the first order approximation, deviations from MOND are proportional to $\alpha$.
This is typically a small number for galaxies.
To see this, first note that we typically have $\sqrt{f_G}/m \gtrsim \mathrm{Mpc}$ \citep{Skordis2020}.
This is much larger than the MOND radius of galaxies, which typically satisfies $r_{\mathrm{MOND}} \lesssim 10\,\mathrm{kpc}$.
Thus, $\alpha$ is typically smaller than $10^{-4}$ for galaxies,
\begin{align}
 \left.\alpha\right|_{\mathrm{galaxies}} = \frac{m^2}{f_G} r_{\mathrm{MOND}}^2 \lesssim 10^{-4} \ll 1 \,.
\end{align}
In contrast, for galaxy clusters, the MOND radius can be large enough to give $\alpha = \mathcal{O}(1)$.

\section{Numerical solutions}
\label{sec:appendix:numerical}

In addition to the analytical approximation discussed above, we also made use of numerical solutions.
To obtain these, we used the Julia package `OrdinaryDiffEq.jl` with the `AutoTsit5(Rosenbrock23())` method \citep{Rackauckas2017, Tsitouras2011}.

We again used the splits $\hat{\Phi} = \hat{\Phi}_b + \hat{\Phi}_c$ and $\varphi = \varphi_b + \varphi_c$ and numerically solved for $\hat{\Phi}_c$ and $\varphi_c$.
The equations of motion are
\begin{align}
\label{eq:numericaleom}
&\hat{\Phi}_c'' + \frac{2 \hat{\Phi}_c'}{r} = S \\
&\begin{multlined}[t]\frac{\varphi_c'}{r} + \varphi_c'' \frac{1 + F(\varphi_b + \varphi_c)}{2}
+ \varphi_b'' \frac{F(\varphi_b + \varphi_c) - F(\varphi_b)}{2}
\\= \frac{S}{2 \tilde{\mu}\left(|\varphi_b' + \varphi_c'|/a_0\right)} \,,
\end{multlined}
\end{align}
where
\begin{align}
 S &\equiv m^2 \left(\mu/Q_0 - \varphi_b - \varphi_c - \hat{\Phi}_b - \hat{\Phi}_c \right) \,, \\
 \tilde{\mu}(s) &=  f_G \frac{
   1 + 2s (1 - f_G) - \sqrt{1 + 4s(1 - f_G)}
 }{
   2s(1-f_G)^2
 } \,, \\
 F(\varphi) &\equiv \frac{\tilde{\mu}'\left(\frac{|\vec{\nabla} \varphi|}{a_0}\right)\frac{|\vec{\nabla} \varphi|}{a_0}}{\tilde{\mu}\left(\frac{|\vec{\nabla} \varphi|}{a_0}\right)} = \frac{1}{\sqrt{1 + 4 (1- f_G)\frac{|\vec{\nabla} \varphi|}{a_0} }} \,.
\end{align}
On the right-hand side of the $\varphi_c$ equation, there is in general an additional term proportional to
\begin{align}
\label{eq:appendix:numerical:leftout}
4 \pi G_N f_G \rho_b \left(
  \frac{1}{2 \tilde{\mu}(|\varphi_b' + \varphi_c'|/a_0)} -
  \frac{1}{2 \tilde{\mu}(\varphi_b'/a_0)}\right) \,.
\end{align}
We left out this term since it vanishes for a baryonic point particle, $\rho_b = M_b \delta(\vec{x})$, which is what we consider here.
To see that it vanishes, we first note that it vanishes outside $r = 0$ because of the factor $\rho_b$.
For $r\to 0$, we have $\rho_b \propto \delta(r)/r^2$, $\varphi_b' \propto 1/r$, and $\varphi_c' \propto r$.
The behavior of $\varphi_c'$ can, for example, be read off from our approximate analytical solution which is valid at $r \to 0$.
Thus, we can expand $\tilde{\mu}((\varphi_b' + \varphi_c')/a_0)$ around $\varphi_c' = 0$ and Eq.~\eqref{eq:appendix:numerical:leftout} becomes proportional to
\begin{align}
\rho_b \frac{\tilde{\mu}'(\varphi_b'/a_0)}{\tilde{\mu}(\varphi_b'/a_0)^2} \varphi_c' \to (1/r^2) \delta(r) (1/r)^{-3/2} \cdot r = \sqrt{r} \delta(r) = 0 \,,
\end{align}
where we used that $\tilde{\mu}'(s)$ scales as $s^{-3/2}$ at large $s$ while $\tilde{\mu}(s)$ becomes constant.

We used a dimensionless length variable $y = r/l$ with $l = 1\,\mathrm{kpc}$.
We solved the equations in terms of $v$ and $u$, which are obtained from $\varphi_c$ and $\hat{\Phi}_c$, respectively, by rescaling and absorbing the constant $\mu/Q_0$,
\begin{align}
 \mu/Q_0 - \varphi_c - \hat{\Phi}_c \equiv - A (u + v) \,,
\end{align}
with $A \equiv 10^{-7}$.
We used the boundary conditions
\begin{align}
 u'(0) = v'(0) &= 0 \,, \\
 u(0) + v(0) &= \mathrm{const} \,.
\end{align}
The first follows from spherical symmetry, the second corresponds to a choice of chemical potential for the ghost condensate.
When comparing the numerical solution to our analytical approximation, the following relation is useful,
\begin{align}
 u(0) + v(0) = - A^{-1} \sqrt{G_N M_b a_0} (p + \ln(r_{\mathrm{MOND}}/l)) \,.
\end{align}
The logarithm accounts for the fact that $p$ is defined for $l = r_{\mathrm{MOND}}$ while we used $l = 1\,\mathrm{kpc}$ for the numerical solution.
To avoid numerical complications from the $1/r$ factor in the equations of motion, we did not impose these boundary conditions at $r=0$ but at $r = 10^{-5}\,\mathrm{kpc}$, corresponding to $y = 10^{-5}$.

\section{Oscillations, $m^2$-$f_G$-degeneracy}
\label{sec:oscillations}

\begin{figure}
 \centering
 \includegraphics[width=\hsize]{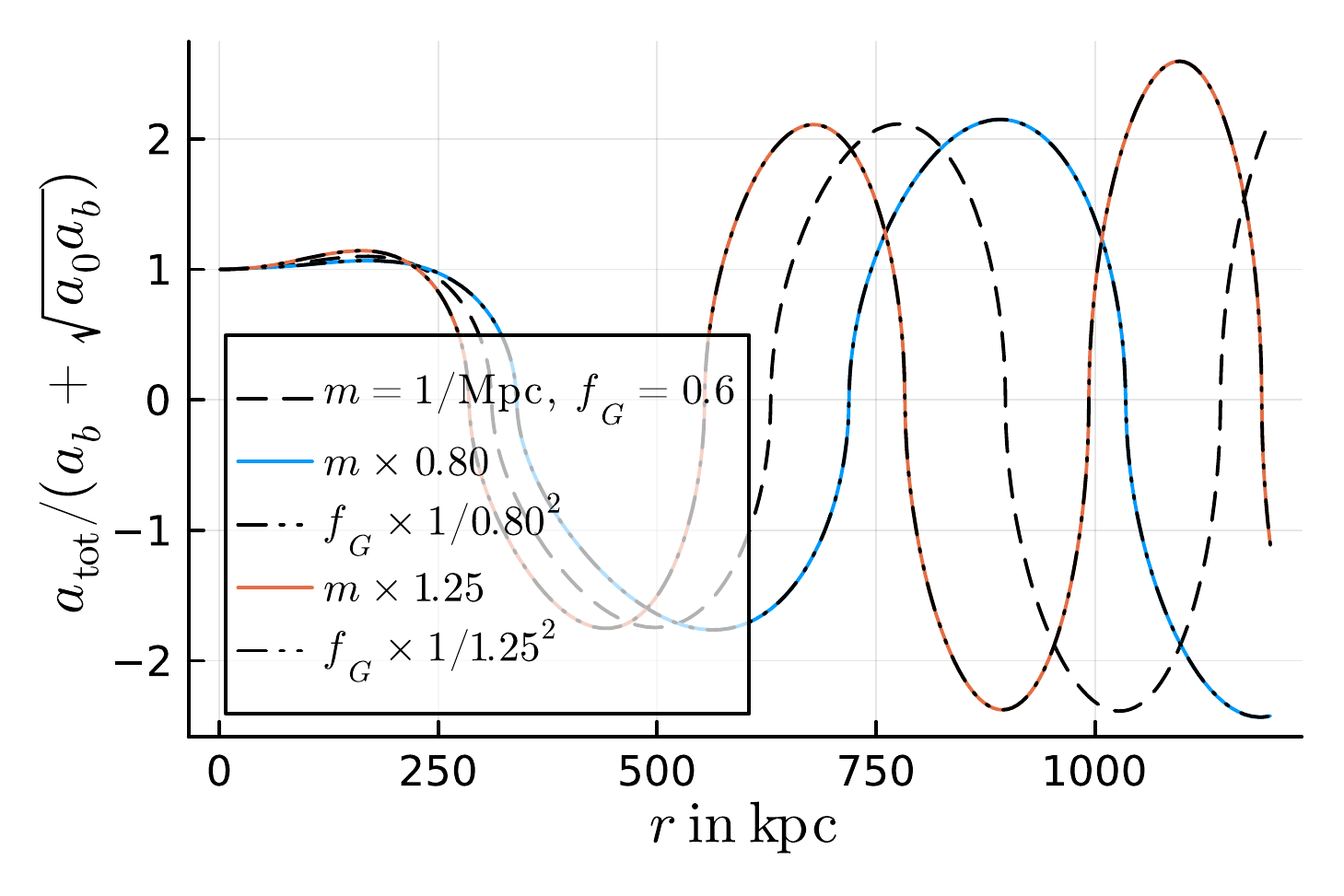}
 \caption{
   Total acceleration $a_{\mathrm{tot}}$ relative to the MOND-like acceleration $a_b + \sqrt{a_0 a_b}$ for a galaxy with $M_b = 2 \cdot 10^{10}\,M_\odot$ for different model parameters $m^2$ and $f_G$ but with the same boundary condition imposed at $r = 0$.
   This boundary condition is chosen such that the first maximum of $a_{\mathrm{tot}}/(a_b + \sqrt{a_0 a_b})$ is $1.1$ for $m = 1\,\mathrm{Mpc}^{-1}$ and $f_G = 0.6$.
   This illustrates that the total acceleration begins to oscillate at large radii.
   It also illustrates that the total acceleration depends only on the combination $m^2/f_G$ but not on $m^2$ and $f_G$ individually.
 }
 \label{fig:illustrate-oscillations-degeneracy}
\end{figure}

Above, we saw that the effective mass $M_{\mathrm{eff}}$ -- and thus also the acceleration -- drops to zero at a finite radius.
Beyond this radius, the acceleration begins to oscillate \citep{Skordis2020, Arkani-Hamed2007}, see for example Fig.~\ref{fig:illustrate-oscillations-degeneracy}.
As discussed above, this oscillatory regime is potentially unstable since it involved negative energy densities and even negative masses.
In this Appendix, we ignore this, keeping in mind that the oscillations might not be physical.

The oscillations depend strongly on the combination $m^2/f_G$ which multiplies the condensate density.
This is illustrated in Fig.~\ref{fig:illustrate-oscillations-degeneracy} which shows multiple solutions with the same boundary condition and the same mass $M_b$ but different values of the parameter $m^2/f_G$.

Indeed, in spherical symmetry, the total acceleration in the AeST model depends not on $m$ and $f_G$ separately but only on the combination $m^2/f_G$.
This is also illustrated in Fig.~\ref{fig:illustrate-oscillations-degeneracy} and can be understood from the integro-differential equation Eq.~\eqref{eq:atotnunaive} that determines the total acceleration.
Crucially, Eq.~\eqref{eq:atotnunaive} depends only on the sum $\hat{\Phi} + \varphi$ but not on $\hat{\Phi}$ and $\varphi$ separately.

\section{Maximum radius of MOND-like behavior, optimal boundary conditions, relation to maximum of $M_{\mathrm{eff}}$}
\label{sec:optimal}

The AeST model deviates from MOND at large radii.
Here, we discuss quantitatively at which radii these deviations occur and how large they are.

\begin{figure}
 \centering
 \includegraphics[width=\hsize]{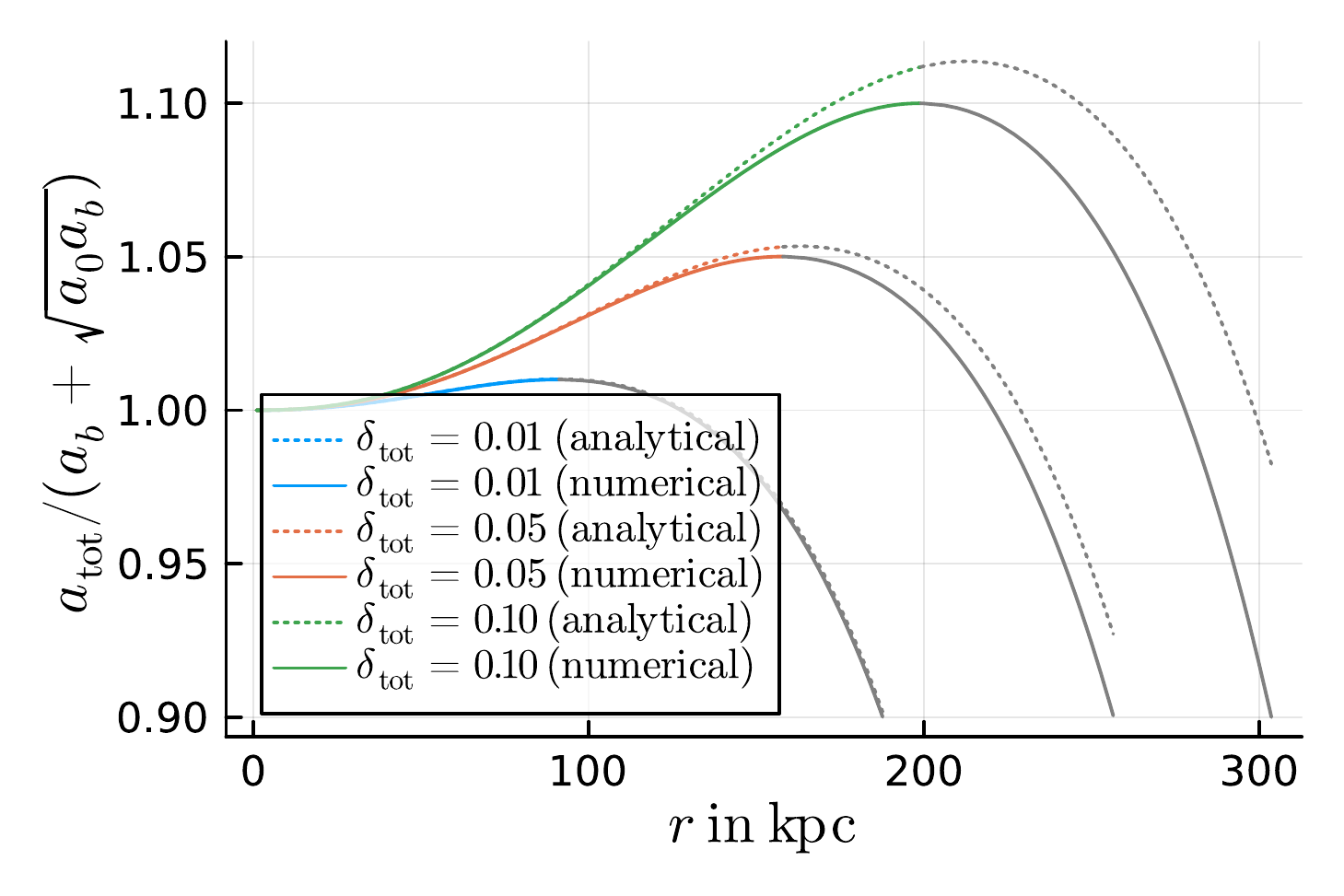}
 \caption{
   Total acceleration $a_{\mathrm{tot}}$ relative to the MOND-like acceleration $a_b + \sqrt{a_0 a_b}$ for numerical (solid lines) and analytical (dotted lines) solutions for various boundary conditions for a galaxy with $M_b = 2 \cdot 10^{10}\,M_\odot$ and $f_G/m^2 = 0.99\,\mathrm{Mpc}^2$.
   The boundary conditions are chosen such that the maxima of $a_{\mathrm{tot}}/(a_b + \sqrt{a_0 a_b})$ for the numerical solutions are $1 + \delta_{\mathrm{tot}} = 1.01$ (green), $1.05$ (red), and $1.10$ (blue).
   For each solution, we indicate the region where the condensate density of the numerical solution is negative with gray line colors.
 }
 \label{fig:illustrate-bcdependence}
\end{figure}

First, we point out that deviations from MOND are inevitable.
By this we mean that, for given model parameters and a given baryonic mass $M_b$, one cannot push the deviations to arbitrarily large radii by judiciously adjusting the boundary condition.
This can be seen by considering the effective mass $M_{\mathrm{eff}}$.
If deviations from MOND are small, the zeroth order approximation $M_{\mathrm{eff},0} = M_b$ should be close to our first order approximation $M_{\mathrm{eff},1}$,
\begin{align}
 \frac{M_{\mathrm{eff},1}(r)}{M_b} =1 + \alpha x^2 \left(\frac12 + \frac19 x \left(3 \, p + 1 - 3 \ln(x) \right) \right)\,,
\end{align}
where, again, $p$ parametrizes the boundary condition, $x = r/r_{\mathrm{MOND}}$, and $\alpha = m^2 r_{\mathrm{MOND}}^2/f_G$.
The general form of this function is that it approaches one for $r \to 0$,
  has a maximum at some finite radius, and then drops to zero (see e.g., Fig.~\ref{fig:illustrate-bcdependence}).
This definitely deviates from the zeroth order approximation (i.e., from MOND) when it drops to zero.
And we can obviously make the radius where this happens arbitrarily large by making the boundary condition $p$ arbitrarily large.
So one might wonder whether we can avoid deviations from MOND by just making $p$ arbitrarily large.
But this is not possible because the larger $p$ is, the more $M_{\mathrm{eff}}$ deviates from MOND at its maximum.
So there must be a finite optimal boundary condition $p$.
For this optimal boundary condition, the radius up to which deviations from MOND stay small is maximized.
We refer to this radius as $r_{\mathrm{max}}$.

More concretely, we considered the ratio of the acceleration $a_{\tilde{\varphi}}$ and its value without deviations from MOND, that is, we considered $a_{\tilde{\varphi}}/\sqrt{a_0 a_b}$.
If one allows this ratio to deviate from MOND by at most a fraction $\delta$,
\begin{align}
 \left| \frac{a_{\tilde{\varphi}}(r)}{\sqrt{a_0 a_b(r)}} -1 \right| = \left|\mathrm{sign}(M_{\mathrm{eff}}(r)) \sqrt{\frac{|M_{\mathrm{eff}}(r)|}{M_b}} -1 \right| \stackrel{!}{<} \delta \,.
\end{align}
That is, if one allows $\sqrt{M_{\mathrm{eff}}/M_b}$ to deviate from $1$ by at most a fraction $\delta$.
Then, there is an optimal boundary condition $p$ that allows this condition to be fulfilled up to a maximum possible radius $r_{\mathrm{max}}$.

As we show in Appendix~\ref{sec:approx:optimal-bc}, this optimal boundary condition is that for which $\sqrt{M_{\mathrm{eff}}/M_b} = a_{\tilde{\varphi}}/\sqrt{a_0 a_b}$ has the value $1+\delta$ at its maximum.
We refer to this radius where $a_{\tilde{\varphi}}$ reaches its maximum as $r_{\mathrm{maxratio}}$.
We illustrate the meaning of the quantities $r_{\mathrm{max}}$, $r_{\mathrm{maxratio}}$, and $\delta$ in Fig.~\ref{fig:illustrate-bestIC-terminology}.

Instead of $a_{\tilde{\varphi}}$, we can also consider the total acceleration $a_{\mathrm{tot}} = a_{\tilde{\varphi}} + a_{\tilde{\Phi}}$.
This gives an analogous result:
If we allow the total acceleration $a_{\mathrm{tot}}$ to deviate from MOND by at most a fraction $\delta_{\mathrm{tot}}$, then the optimal boundary condition is that for which $a_{\mathrm{tot}}/(a_b + \sqrt{a_0 a_b})$ has the value $1+\delta_{\mathrm{tot}}$ at its maximum.

Strictly speaking, the optimal boundary conditions for $a_{\tilde{\varphi}}/\sqrt{a_0 a_b}$ and $a_{\mathrm{tot}}/(a_b + \sqrt{a_0 a_b})$ differ.
At least for galaxies, however, the optimal boundary conditions are almost identical for both cases.
Moreover, the maximizing radius and the value at this maximum are almost identical.
This is shown in Appendix~\ref{sec:approx:maximum}.
Thus, for our purposes, it does not matter much whether we consider the optimal boundary condition for $a_{\tilde{\varphi}}$ or for $a_{\mathrm{tot}}$.

\begin{figure}
 \centering
 \includegraphics[width=\hsize]{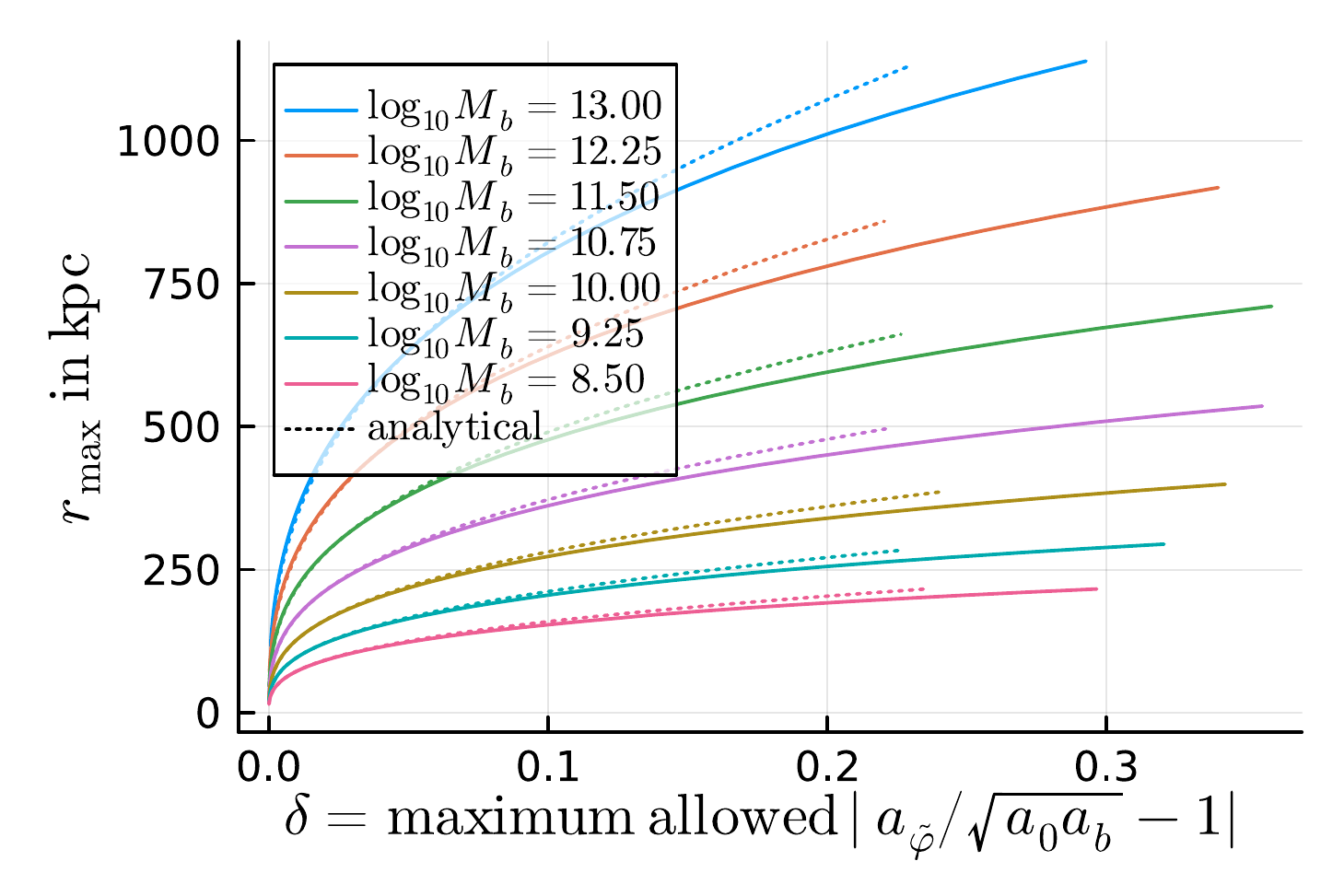}
 \caption{
   Radius $r_{\mathrm{max}}$ up to which the acceleration $a_{\tilde{\varphi}} = \sqrt{a_0 a_b} \sqrt{M_{\mathrm{eff}}/M_b}$ can, at best, stay within a fraction $\delta$ of the MOND-like acceleration $\sqrt{a_0 a_b}$ as a function of $\delta$.
   This corresponds to the radius where $a_{\tilde{\varphi}}/\sqrt{a_0 a_b} = 1-\delta$ for the optimal boundary condition.
   This is for $f_G/m^2 = 0.99\,\mathrm{Mpc}^2$.
   We show the result for both analytical (solid lines) and analytical (dotted lines) solutions and for various baryonic masses $M_b$.
   Results for the analytical approximation are shown only where our estimate Eq.~\eqref{eq:validity} says that the approximation is better than $q=10\%$.
   For the analytical solution we further assume $r_{\mathrm{max}} = 1.53 \, r_{\mathrm{maxratio}}$.
 }
 \label{fig:illustrate-delta-rmax}
\end{figure}

\begin{figure}
 \centering
 \includegraphics[width=\hsize]{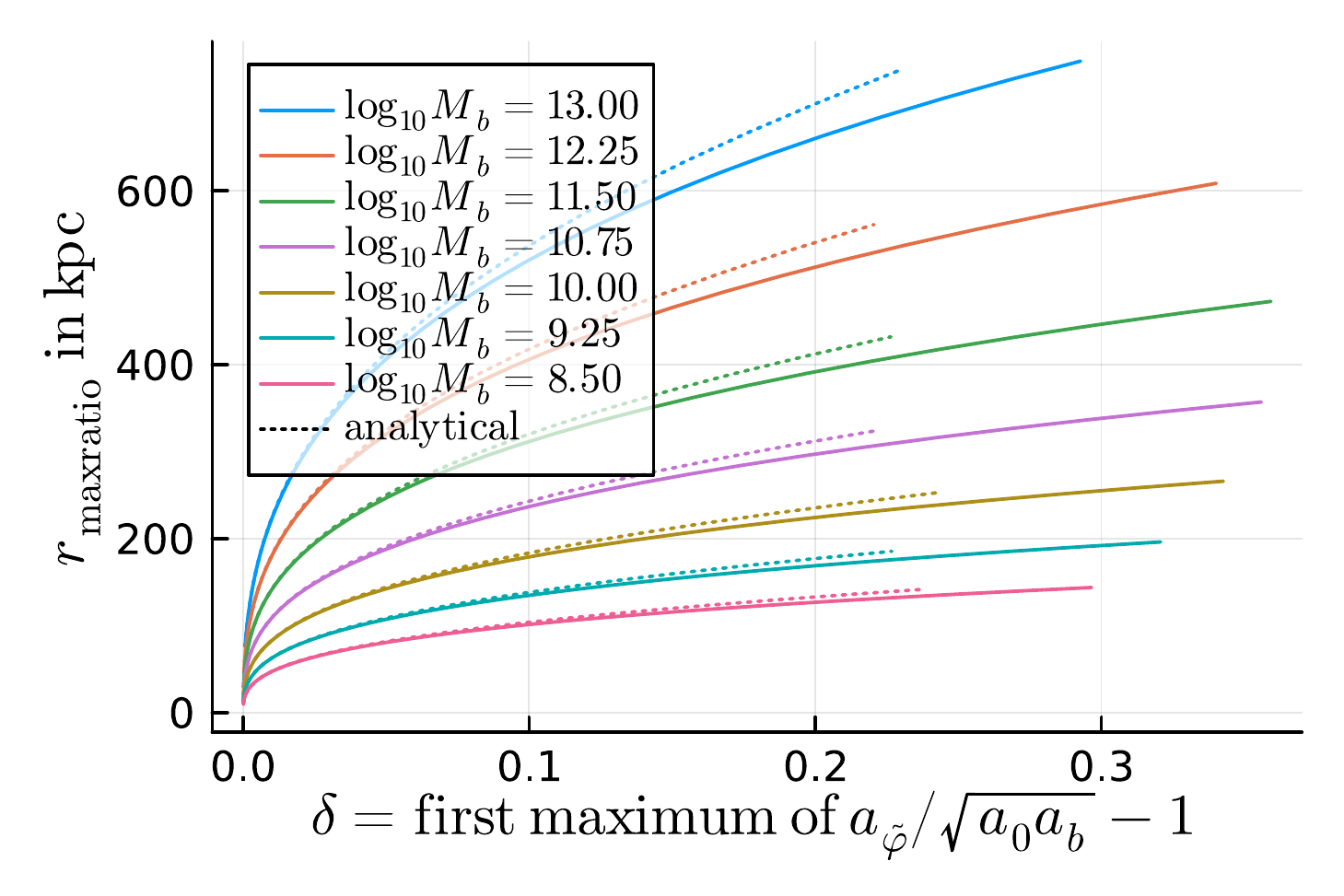}
 \caption{
   Radius $r_{\mathrm{maxratio}}$ where the first maximum of the ratio $a_{\tilde{\varphi}} / \sqrt{a_0 a_b}$ occurs for boundary conditions where this maximum deviates from MOND by a fraction $\delta$.
   These boundary conditions are the optimal boundary conditions given a maximum allowed deviation $\delta$ of $a_{\tilde{\varphi}}$ from MOND.
   This is for $f_G/m^2 = 0.99\,\mathrm{Mpc}^2$.
   We show the result for both analytical (solid lines) and analytical (dotted lines) solutions and for various baryonic masses $M_b$.
   Results for the analytical approximation are shown only where our estimate Eq.~\eqref{eq:validity} says that the approximation is better than $q=10\%$.
 }
 \label{fig:illustrate-delta-rmaxratio}
\end{figure}

In Fig.~\ref{fig:illustrate-delta-rmax} we show the relation between the maximum allowed deviation from MOND $\delta$ and the maximum radius $r_{\mathrm{max}}$ up to which this condition can be fulfilled.
For the numerical solutions, we found $r_{\mathrm{max}}$ by maximizing the radius where a solution first deviates by more than a fraction $\delta$ from MOND using the Julia package `Optim.jl` \citep{Mogensen2018}.

As discussed above, for the optimal boundary conditions, the ratio  $a_{\tilde{\varphi}}/\sqrt{a_0 a_b}$ has the value $1+\delta$ at its first maximum.
The radius $r_{\mathrm{maxratio}}$ where this maximum occurs is closely related to the radius $r_{\mathrm{max}}$.
This is illustrated in Fig~\ref{fig:illustrate-delta-rmaxratio} which shows $r_{\mathrm{maxratio}}$ as a function of $\delta$.
This is very similar to Fig.~\ref{fig:illustrate-delta-rmax} just with the values on the y-axis a bit larger.
In particular, $r_{\mathrm{maxratio}}$ corresponds to $a_{\tilde{\varphi}}/\sqrt{a_0 a_b} = 1 + \delta$, while $r_{\mathrm{max}}$ corresponds to $a_{\tilde{\varphi}}/\sqrt{a_0 a_b} = 1 - \delta$.

In Appendix~\ref{sec:approx:maximum} and Appendix~\ref{sec:approx:rmax-vs-rmaxratio}, we give an analytical estimate for the relation between $\delta$, $r_{\mathrm{max}}$, and $r_{\mathrm{maxratio}}$.
For galaxies, a good approximation is
\begin{align}
 \frac{r_{\mathrm{maxratio}}}{r_{\mathrm{MOND}}} \approx
 \left(9 \frac{(1+\delta)^2-1}{r_{\mathrm{MOND}}^2 \,m^2/f_G}\right)^{1/3} \,, \quad
 r_{\mathrm{max}} \approx 1.53 \, r_{\mathrm{maxratio}} \,.
\end{align}
Thus, where this approximation is valid, the only difference between Fig.~\ref{fig:illustrate-delta-rmaxratio} and Fig.~\ref{fig:illustrate-delta-rmax} is a factor $1.53$ in the y-axis values.

Our estimate for $r_{\mathrm{max}}$ can be compared to a related estimate from \citet{Skordis2020}.
There, the authors estimate that the AeST model acceleration is MOND-like up to a critical radius $r_C$,
\begin{align}
 \frac{r_C}{r_{\mathrm{MOND}}} \sim \left(\frac{1}{m^2 \, r_{\mathrm{MOND}}^2}\right)^{1/3} \,.
\end{align}
This has the same scaling in $r_{\mathrm{MOND}}$ and $m$ as our estimate.
However, our estimate improves on this in a few ways.
First, the $m^2$ factor in $r_C$ should be $m^2/f_G$.
Otherwise, the estimate does not correctly take into account the difference between $G_N$ and $\hat{G}$.
Second, our version comes with worked out prefactors, including a parameter that controls how big deviations actually are.

We emphasize again that the maximum radius $r_{\mathrm{max}}$ -- or the critical radius $r_C$ -- corresponds to choosing boundary conditions that are optimal for reproducing MOND (given an allowed deviation $\delta$).
Galaxies are not guaranteed to reproduce MOND up to that radius.
In general, galaxies will end up with boundary conditions different from the optimal one and deviate from MOND already at smaller radii.

In Fig.~\ref{fig:illustrate-delta-rmax} and Fig.~\ref{fig:illustrate-delta-rmaxratio},
  we show our analytical approximation only where we analytically estimate that it deviates by at most $q=10\%$ from the full solution.
This validity estimate works by going to the next-higher order in our method for approximating $M_{\mathrm{eff}}$ and then comparing to our first-order approximation, see Appendix~\ref{sec:approx:validity}.
Essentially, our approximation is valid for small deviations from MOND $\delta$, but not for larger ones, as can already be guessed from Fig.~\ref{fig:illustrate-bcdependence}.

From Fig.~\ref{fig:illustrate-delta-rmax} and Fig.~\ref{fig:illustrate-delta-rmaxratio} we see that more massive galaxies can remain close to MOND for longer than less massive galaxies.
However, in the context of MOND, accelerations are often more important than radii.
Thus, in Fig.~\ref{fig:illustrate-delta-abmin}, we show the minimum acceleration $a_{b,\mathrm{min}} = G_N M_b/r_{\mathrm{max}}^2$ corresponding to the maximum radius $r_{\mathrm{max}}$.
We see that the trend is now inverted due to the additional factor of $M_b$ in $a_b$.
Less massive galaxies can have MOND-like behavior down to smaller accelerations than more massive galaxies.

\section{Stacking for weak lensing}
\label{sec:appendix:stacking}

In the weak-lensing analysis of \citet{Brouwer2021}, the central quantity is the stacked excess surface density (ESD) profile $\Delta \Sigma_{\mathrm{stacked}}$.
Stacking here means taking a weighted average over the galaxy sample.
More specifically,
\begin{align}
\Delta \Sigma_{\mathrm{stacked}} = \frac{\Sigma_i W_i \left( \frac{1}{1+\mu} \epsilon_{t,i} \Sigma_{\mathrm{crit},i}\right)}{\Sigma_i W_i} \,,
\end{align}
where the $W_i$ are weights, $\Sigma_{\mathrm{crit}}$ is the critical surface density, $\epsilon_t$ is the ellipticity, and the factor $1+\mu$ calibrates the shear estimates.
The ellipticity $\epsilon_t$ of a galaxy is the sum of its intrinsic ellipticity $\epsilon_t^{\mathrm{int}}$ and the tangential shear $\gamma_t$ caused by weak lensing.
The intrinsic ellipticities average to zero in a large sample, so that $\Delta \Sigma_{\mathrm{stacked}}$ measures the tangential shear $\gamma_t$.
For an individual lens and up to the calibration factor $1+\mu$, the combination $\gamma_t \Sigma_{\mathrm{crit}}$ is given by the ESD $\Delta \Sigma$,
\begin{align}
\Delta \Sigma(R) = \frac{2 \pi \int_0^R dR' \Sigma(R')}{\pi R^2} - \Sigma(R) \,,
\end{align}
where $\Sigma(R)$ is the surface density corresponding to the lensing mass $M_{\mathrm{lens}}$.
Thus, for a galaxy sample with known surface densities, we can calculate $\Delta \Sigma_{\mathrm{stacked}}$ by simply averaging the individual ESD profiles of each galaxy,
\begin{align}
 \Delta \Sigma_{\mathrm{stacked}} = \frac{\Sigma_i W_i \Delta \Sigma_i}{\Sigma_i W_i} \,.
\end{align}

The stacked ESD profile is linear in the surface densities $\Sigma_i$.
These surface densities are calculated linearly from the density $\rho_{\mathrm{lens}}$ that produces the lensing mass $M_{\mathrm{lens}}$.
This lensing mass is defined by $\Phi'(r) = G_N M_{\mathrm{lens}}/r^2$.
In our case,
\begin{align}
 \rho_{\mathrm{lens}}(r) = \frac{1}{4\pi G_N \, r^2} \partial_r \left(r^2 \partial_r(\hat{\Phi} + \varphi)\right) \,.
\end{align}
Thus, the stacked ESD profile is linear in $\hat{\Phi} + \varphi$.
As a consequence, the total acceleration $a_{\mathrm{tot}}$ inferred from the stacked ESD profile is just the weighted average of the total accelerations $a_{\mathrm{tot},i}$ of each stacked galaxy,
\begin{align}
 a_{\mathrm{tot},\mathrm{stacked}}(R) = \frac{\Sigma_i W_i \, a_{\mathrm{tot},i}(R)}{\Sigma_i W_i} \,.
\end{align}
In order to calculate a stacked RAR, we should not stack at a fixed position $R$ but at a fixed acceleration $a_b = G_N M_b/R^2$.
So we should instead use
\begin{align}
 a_{\mathrm{tot},\mathrm{stacked}}(a_b) = \frac{\Sigma_i W_i \, a_{\mathrm{tot},i}\left(\sqrt{\frac{G_N M_{b,i}}{a_b}}\right)}{\Sigma_i W_i} \,.
\end{align}
For a fixed baryonic mass $M_b$, stacking in position space and acceleration space is equivalent.

We now illustrate how a stacked weak-lensing RAR in the AeST model might look like.
For simplicity, we assume that all weights $W_i$ are the same and we further assume that all galaxies have the same baryonic mass $M_b$.
Essentially, we consider stacking a sample of galaxies that differ only in their boundary conditions.
Then, we have
\begin{align}
 a_{\mathrm{tot},\mathrm{stacked}}(R) = N^{-1}\, \Sigma_i \, a_{\mathrm{tot},i}(R) \,,
\end{align}
where $N$ is the number of galaxies in the sample.

One can also consider a large number of galaxies with boundary conditions $p$ distributed uniformly in an interval $[p_1, p_2]$.
Then, we can write
\begin{align}
 a_{\mathrm{tot},\mathrm{stacked}}(R) = \frac1{p_2 - p_1} \int_{p_1}^{p_2} dp \, a_{\mathrm{tot},p}(R) \,.
\end{align}
As long as all accelerations $a_{\mathrm{tot},p}(R)$ point in the usual direction, that is, as long as $M_{\mathrm{eff}}$ is positive, this integral can be done analytically for our first order analytical approximation,
\begin{multline}
\label{eq:atotstackedanalytical}
\left.a_{\mathrm{tot},\mathrm{stacked}}(R)\right|_{M_{\mathrm{eff}} > 0} = a_b \frac{M_{\mathrm{eff}}(x, \bar{p})}{M_b} \\
+\frac{\sqrt{a_0 a_b}}{p_2 - p_1} \frac{2}{\alpha x^3} \left(
\left(\frac{M_{\mathrm{eff}}(x, p_2)}{M_b}\right)^{3/2} -
\left(\frac{M_{\mathrm{eff}}(x, p_1)}{M_b}\right)^{3/2}
\right) \,,
\end{multline}
where $x = r/r_{\mathrm{MOND}}$, $\bar{p} = (p_1 + p_2)/2$, and $M_{\mathrm{eff}}(x, p)$ is our first order analytical approximation Eq.~\eqref{eq:Meffx}.

\section{Higher-order terms in condensate density}
\label{sec:appendix:higherorder}

The ghost condensate density from Eq.~\eqref{eq:rhoc} is linear in the fields $\varphi$ and $\hat{\Phi}$.
In contrast, for their cosmological calculations, \citet{Skordis2020} used a ghost condensate density including nonlinear corrections.
Here, we discuss whether using the linearized form around galaxies is valid and explain our decision to do so.

In the full action, not assuming the quasi-static weak-field limit, the condensate density corresponds to a term
\begin{align}
 \label{eq:KQquadratic}
 \mathcal{K}(Q) = \mathcal{K}_2 (Q - Q_0)^2 \,,
\end{align}
where $Q = A^\mu \nabla_\mu \phi$ is a scalar combination of the normalized vector field $A^\mu$ and the scalar field $\phi$.
In the quasi-static weak-field limit, $\phi = Q_0 \cdot t + \varphi$.
The constant $\mathcal{K}_2$ is related to the mass parameter $m$ by $m = \sqrt{2\mathcal{K}_2/(2-K_B)} Q_0$.

Our expression for the ghost condensate Eq.~\eqref{eq:rhoc} is linear because the function $\mathcal{K}(Q)$ is quadratic.
Indeed, in the quasi-static weak-field limit,
\begin{align}
 Q - Q_0 \approx Q_0 \cdot \left(\frac{\dot{\varphi}}{Q_0} - \varphi - \hat{\Phi}\right) \,.
\end{align}

\begin{figure}
 \includegraphics[width=\hsize]{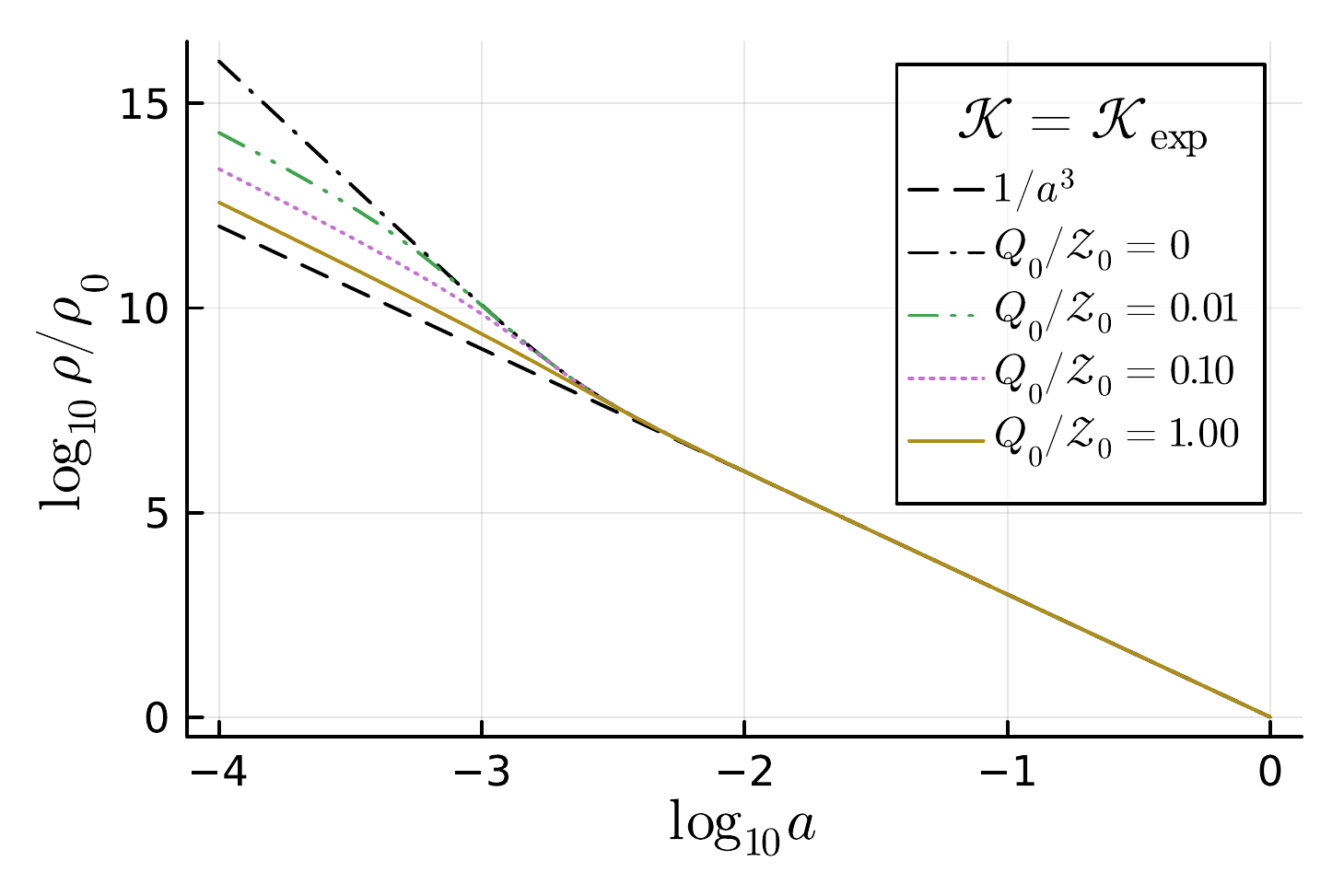}
 \caption{
   Cosmological ghost condensate density $\rho$ as a function of the scale factor $a$ relative to its density $\rho_0$ today, at $a=1$, for various values of the combination $Q_0/\Zcal_0$.
   This is for $m = 1\,\mathrm{Mpc}^{-1}$, $K_B = 0.1$, $H_0 = 70\,\mathrm{km}\,\mathrm{s}^{-1}\,\mathrm{Mpc}^{-1}$, and $\Omega_0 = 0.25$.
   The ghost condensate follows a dust-like evolution only at late times.
   Larger values of $m$ allow for dust-like evolution even at $a = 10^{-4}$ but are in conflict with having MOND-like behavior around galaxies.
 }
 \label{fig:cosmo-rho-vs-a-exp}
\end{figure}

However, a quadratic function $\mathcal{K}(Q)$ cannot simultaneously satisfy cosmological constraints and support a MOND-like regime in galaxies \citep{Skordis2020}.
In particular, cosmological observations require that the ghost condensate's equation of state $w$ satisfies $w \lesssim 0.02$ at a scale factor $a = 10^{-4}$ \citep{Ilic2021}.
But if we assume $m^2/f_G \lesssim 1/\mathrm{Mpc}$ in order to have a MOND-like phenomenology around galaxies, $w$ is too large.
Indeed, using $0 < K_B < 2$ and $f_G < 1$, \citep{Skordis2020}
\begin{align}
w \approx \frac{3 H_0^2 \Omega_0}{2(2-K_B) f_G (m^2/f_G) a^3} \gtrsim 10^{-8} a^{-3} \,,
\end{align}
where $H_0$ is the Hubble constant and $\Omega_0$ is the matter density parameter today.
This is illustrated in Fig.~\ref{fig:cosmo-rho-vs-a-exp} which shows the cosmological ghost condensate density as a function of the scale factor $a$ for $m = 1\,\mathrm{Mpc}^{-1}$ (see the $Q_0/\Zcal_0 = 0$ line, the parameter $\Zcal_0$ is discussed below).
We see that a dust-like evolution is possible at late times but not around $a \sim 10^{-4}$.

To avoid this problem, \citet{Skordis2020} introduced two alternative forms of $\mathcal{K}(Q)$,
\begin{align}
 \mathcal{K}_{\mathrm{exp}}(Q) &=  \mathcal{K}_2 \Zcal_0^2 \left(e^{\Zcal^2}-1\right) \,, \\
 \mathcal{K}_{\mathrm{cosh}}(Q) &= 2 \mathcal{K}_2 \Zcal_0^2 \left(\cosh(\Zcal)-1\right) \,,
\end{align}
where $\Zcal_0$ is a constant and
\begin{align}
 \Zcal \equiv \frac{Q-Q_0}{\Zcal_0}\,.
\end{align}
These suppress the equation of state at early times where $\Zcal$ is large and reduce to the quadratic $\mathcal{K}(Q)$ at small $\Zcal$.

In our galaxy-scale calculations above, we assumed the quadratic form of $\mathcal{K}(Q)$.
This is justified only if $\Zcal$ is sufficiently small.
To see whether or not this is the case, we first note that a typical value of $\dot{\varphi}/Q_0 - \varphi - \hat{\Phi}$ around galaxies is $10^{-7}$.
Thus, a typical value of $\Zcal$ around galaxies is $\Zcal \sim \frac{Q_0}{\Zcal_0} \cdot 10^{-7}$.
That is, we can use the quadratic $\mathcal{K}(Q)$ as long as
\begin{align}
 \frac{Q_0}{\Zcal_0} \lesssim 10^7 \,.
\end{align}

Interestingly, this condition is not satisfied, or only barely satisfied, for the explicit cosmological perturbation calculations in \citet{Skordis2020}.
In particular, \citet{Skordis2020} used $Q_0/\Zcal_0 = 10^8$ for their example calculation using $\mathcal{K}_{\mathrm{cosh}}$ and $Q_0/\Zcal_0 = 10^{13}$ for $\mathcal{K}_{\mathrm{exp}}$.

\begin{figure}
 \includegraphics[width=\hsize]{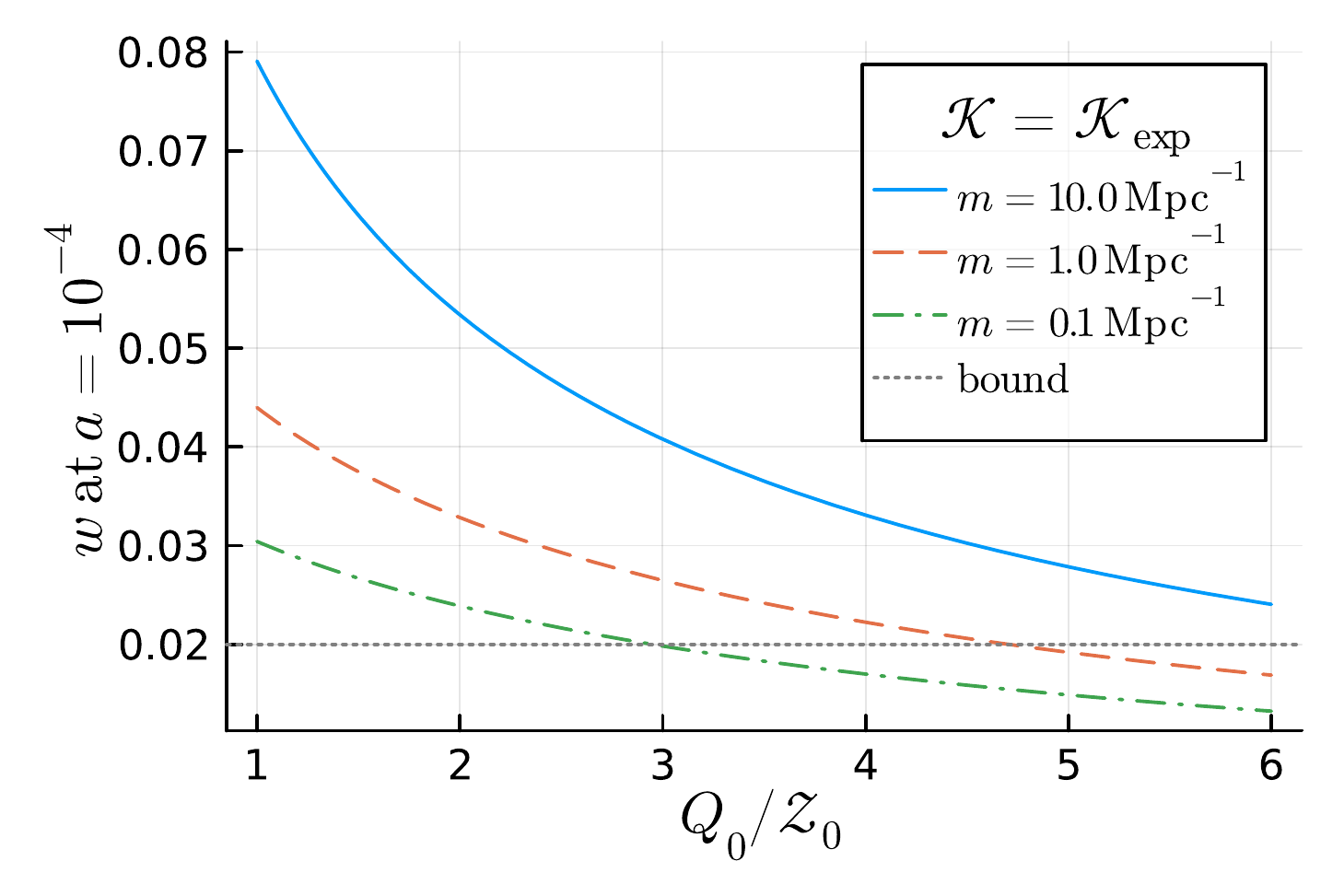}
 \caption{
   Equation of state at $a=10^{-4}$ as a function of $Q_0/\Zcal_0$ for the exponential $\mathcal{K}(Q)$ function for various masses $m$.
   This is for $K_B = 0.1$, $H_0 = 70\,\mathrm{km}\,\mathrm{s}^{-1}\,\mathrm{Mpc}^{-1}$, and $\Omega_0 = 0.25$.
   The dotted gray line shows the upper bound from \citet{Ilic2021}.
   We note that $m^2$ is the prefactor of $\mathcal{K}(Q)$.
   Thus, both the density and the pressure are proportional to $m^2$ and this prefactor cancels in $w$.
   The dependence on $m$ shown here comes from the constraint that the density parameter today is $\Omega_0$.
 }
 \label{fig:cosmo-w-vs-Q0Z0-exp}
\end{figure}

\begin{figure}
 \includegraphics[width=\hsize]{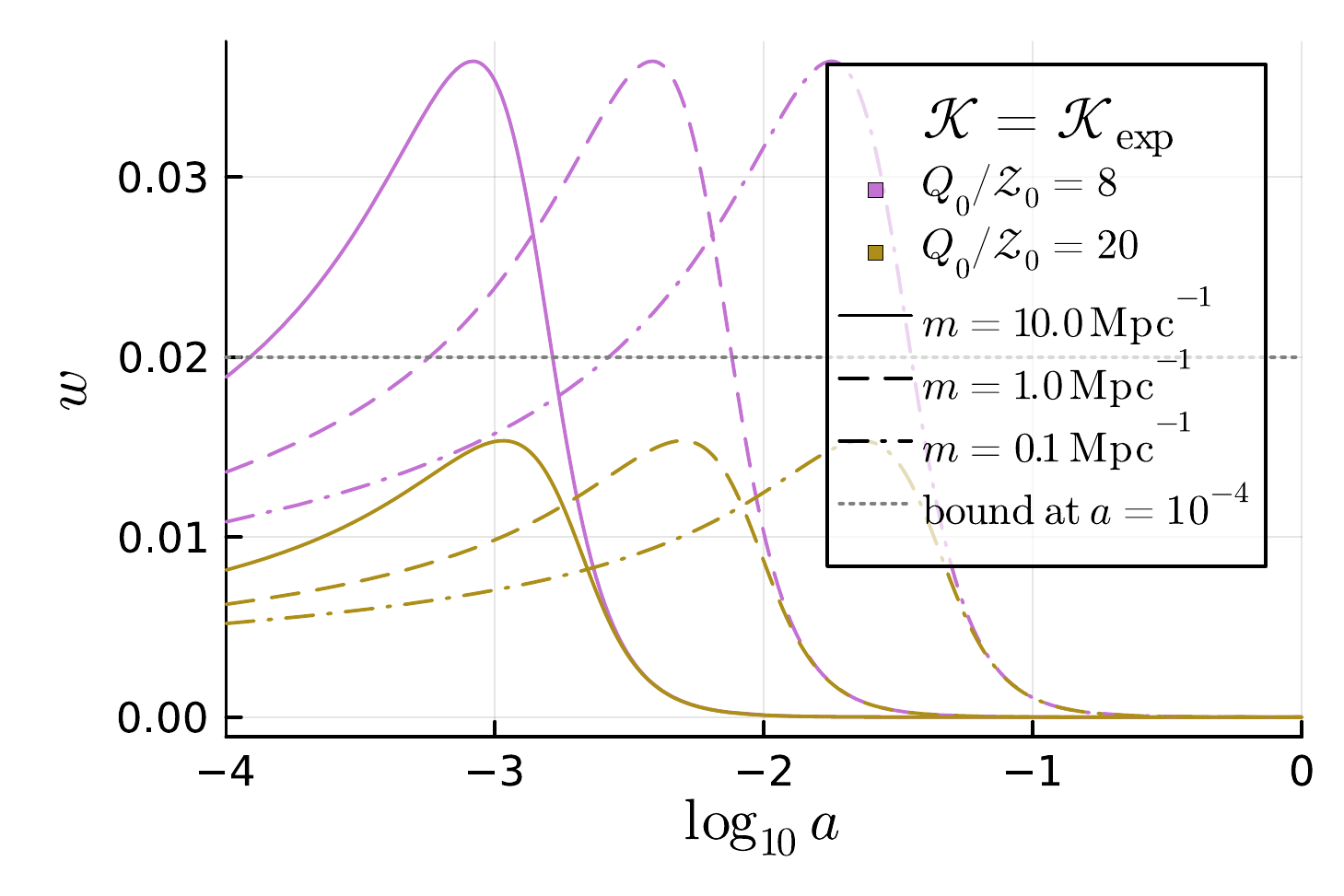}
 \caption{
   Equation of state at as a function of the scale factor $a$ for the exponential $\mathcal{K}(Q)$ function for various masses $m$ and ratios $Q_0/\Zcal_0$.
   This is for $K_B = 0.1$, $H_0 = 70\,\mathrm{km}\,\mathrm{s}^{-1}\,\mathrm{Mpc}^{-1}$, and $\Omega_0 = 0.25$.
   The dotted gray line shows the upper bound at $a = 10^{-4}$ from \citet{Ilic2021}.
 }
 \label{fig:cosmo-w-vs-a-exp}
\end{figure}

We nevertheless assume the quadratic form around galaxies for the following reasons.
First, this seems to be what the authors of the model had in mind.
After all, they use the quadratic form of $\mathcal{K}(Q)$ when deriving the action for the quasi-static weak-field limit \citep{Skordis2020}.

Second, large values of $Q_0/\Zcal_0$ do not seem to be required in order to satisfy cosmological constraints.
Indeed, the constraint $w \lesssim 0.02$ at $a = 10^{-4}$ can easily be satisfied with a much smaller value of $Q_0/\Zcal_0$.
This is illustrated for $\mathcal{K}_{\mathrm{exp}}$ in Fig.~\ref{fig:cosmo-w-vs-Q0Z0-exp} which shows that $Q_0/\Zcal_0 = \mathcal{O}(1)$ is sufficient.

Even if $w$ satisfies $w \lesssim 0.02$ at $a = 10^{-4}$, it can be larger at later times and potentially violate observational constraints that apply at these later times.
Indeed, $w(a)$ is not a monotonous function, as is illustrated in Fig.~\ref{fig:cosmo-w-vs-a-exp}.
Its maximum is set by $Q_0/\Zcal_0$ and $m$ controls at which scale factor $a$ this maximum occurs.
Still, we see from Fig.~\ref{fig:cosmo-w-vs-a-exp} that $Q_0/\Zcal_0 = \mathcal{O}(10)$ is sufficient for $w$ to satisfy $w \lesssim 0.02$ even at later times, thus easily satisfying the constraints at these times from \citet{Ilic2021}.
That is, these constraints can easily be satisfied with $Q_0/\Zcal_0 \ll 10^7$ which allows using the quadratic $\mathcal{K}(Q)$ around galaxies.
We find a similar result for $\mathcal{K}_{\mathrm{cosh}}$.\footnote{
  Of course, in contrast to \citet{Skordis2020}, we did not run a full cosmological perturbations calculation.
  We only considered the constraint on $w$ from the background cosmology.
  It is possible that new constraints arise from the CMB or other observations pertaining to cosmological perturbation theory.
  Here, we assume that this is not the case.
}

A third reason is that, if the nonlinearities of $\mathcal{K}_{\mathrm{exp}}$ or $\mathcal{K}_{\mathrm{cosh}}$ were to be important around galaxies, then either many galaxies would deviate very strongly from MOND or there would be almost no deviations from MOND at all.
Both cases do not require a thorough investigation here.
Strong deviations from MOND are ruled out by observations while pure-MOND predictions are already discussed elsewhere.

It remains to explain why $\mathcal{K}_{\mathrm{exp}}$ and $\mathcal{K}_{\mathrm{cosh}}$ have the effects described in the previous paragraph.
We first consider $\mathcal{K}_{\mathrm{exp}}$ and $\mathcal{K}_{\mathrm{cosh}}$ with the parameter $m$ having roughly the value we assumed above for the quadratic function $\mathcal{K}(Q)$, that is, $m \sim 1\,\mathrm{Mpc}^{-1}$.
Then, as soon as nonlinearities become important, the condensate density is enhanced exponentially compared to the case of a quadratic $\mathcal{K}(Q)$.
Thus, deviations from MOND set in exponentially earlier, that is, the radius $r_{\mathrm{max}}$ is exponentially smaller.
For example, the oscillations discussed in Appendix~\ref{sec:oscillations} start at exponentially smaller radii.
This is in conflict with, for example, the observed RAR which is MOND-like as discussed above.

One way out would be to make the prefactor $m^2$ of $\mathcal{K}_{\mathrm{exp}}$ exponentially smaller to keep the ghost condensate from becoming large.
Indeed, by choosing $m^2$ sufficiently small, we can get rid of any significant deviations from MOND around galaxies.
The predictions of the AeST model in this case are just those of MOND so do not require a special investigation.

A middle ground would be to make $m^2$ exponentially smaller, but by just the right amount so that deviations from MOND set in on roughly galactic scales.
However, the maximum radius $r_{\mathrm{max}}$ would still be exponentially sensitive to the galactic potential $\mu/Q_0 - \varphi - \hat{\Phi}$.
Thus, for galaxies with slightly larger $\mu/Q_0 - \varphi - \hat{\Phi}$, deviations from MOND still set in at an exponentially smaller radius compared to the quadratic $\mathcal{K}(Q)$.
Galaxies with slightly smaller $\mu/Q_0 - \varphi - \hat{\Phi}$ would perfectly follow the MOND predictions up to exponentially large radii.
Thus, a significant fraction of galaxies should still deviate strongly from MOND which is not what is observed.

\section{A few useful properties of $M_{\mathrm{eff},1}(r)$}

In this appendix, we consider a few properties of the analytical first order approximation $M_{\mathrm{eff},1}$ from Eq.~\eqref{eq:Meffx},
\begin{align}
 M_{\mathrm{eff},1}(r) = M_b \left[1 + \alpha x^2 \left(\frac12 + \frac19 x \left(3 \, p + 1 - 3 \ln(x) \right) \right) \right]\,.
\end{align}
We sometimes write $M_{\mathrm{eff}}$ instead of $M_{\mathrm{eff},1}$ for brevity.

\subsection{Increasing functions of boundary condition}
\label{sec:approx:increasing-bc}

We first show that both $a_{\tilde{\varphi}}/\sqrt{a_0 a_b}$ and $a_{\mathrm{tot}}/(a_b + \sqrt{a_0 a_b})$ as well as their spatial derivatives are increasing functions of the boundary condition $p$.
At least as long as $M_{\mathrm{eff}}$ is positive.
This technical result will be useful in Appendix~\ref{sec:approx:optimal-bc}.

We first consider $a_{\tilde{\varphi}}/\sqrt{a_0 a_b} = \sqrt{M_{\mathrm{eff}}/M_b}$ as a function of $p$.
Both $M_{\mathrm{eff}}(r)$ and $M_{\mathrm{eff}}'(r)$ are increasing functions of the boundary condition, that is, of $p$, for all $r$.
This follows from $\alpha$ and $r$ being positive.
We then consider $\sqrt{M_{\mathrm{eff}}(r)}$ and $\partial_r \sqrt{M_{\mathrm{eff}}(r)}$ for radii where $M_{\mathrm{eff}}$ is positive.
In this case, $\sqrt{M_{\mathrm{eff}}(r)}$ is an increasing function of $p$ everywhere because $M_{\mathrm{eff}}$ is,
\begin{align}
 \partial_p \sqrt{M_{\mathrm{eff}}(r)} = \frac{\partial_p M_{\mathrm{eff}}}{2 \sqrt{M_{\mathrm{eff}}(r)}} > 0 \,.
\end{align}
The same holds for the spatial derivative $\partial_r \sqrt{M_{\mathrm{eff}}(r)}$ but it is a bit harder to see.
We find
\begin{align}
\begin{split}
 &\partial_p \partial_x \sqrt{M_{\mathrm{eff}}(r)/M_b}\\
 &= \frac{3 \alpha x^2}{\sqrt{2}} \frac{9 + \alpha x^2(\frac92 + x(3p + 1 -3 \ln(x))) + 9 + \alpha x^2((6-\frac92) + x)}{(18 M_{\mathrm{eff}}(r)/M_b)^{3/2}} \\
 &> \frac{3 \alpha x^2}{\sqrt{2}} \frac{9M_{\mathrm{eff}}(r)/M_b}{(18 M_{\mathrm{eff}}(r)/M_b)^{3/2}} > 0
 \,,
\end{split}
\end{align}
which follows because $r$ and $M_{\mathrm{eff}}$ are positive.

We consider next, again assuming $M_{\mathrm{eff}} > 0$, the function
\begin{align}
\frac{a_{\tilde{\varphi}} + a_{\tilde{\Phi}}}{a_b + \sqrt{a_0 a_b}} = \frac{\frac{M_{\mathrm{eff}}}{M_b} \frac{1}{x} + \sqrt{\frac{M_{\mathrm{eff}}}{M_b}}}{\frac{1}{x} + 1} \equiv \frac{H/x + \sqrt{H}}{1/x + 1} \,.
\end{align}
where we defined $H = M_{\mathrm{eff}}/M_b$.
Since $M_{\mathrm{eff}}$ is an increasing function of the boundary condition $p$,
  the same holds for this function.
The slope of this function with respect to $x$ is also an increasing function of $p$, but this is again a bit harder to see.
We have
\begin{multline}
\partial_p \partial_x \frac{H/x + \sqrt{H}}{1/x + 1}
\\= x^2 \alpha \frac{
  2x H (4+3x) + 4(3+2x)H^{3/2} - x^2(1+x) \partial_x H
}{12 (1+x)^2 H^{3/2}} \,,
\end{multline}
where we already inserted the explicit expressions for $\partial_p H$ and $\partial_p \partial_x H$ due to their simple form.
It now suffices to show positivity of the following expression,
\begin{align}
\begin{split}
 &2 H (4+3x) - x(1+x) \partial_x H \\
 &= 2H (4+3x) -x^2(1+x) \alpha (1 + px - x \ln(x)) \,.
\end{split}
\end{align}
In general, this can be negative.
However, here it cannot be due to our assumption that $M_{\mathrm{eff}}$ is positive.
To see this, we first use the definition of $H$ to rewrite $x(p- \ln(x))$ in terms of $H$,
\begin{align}
x (p - \ln(x)) = -\frac32 - \frac{x}{3} + \frac{3}{\alpha x^2}(H-1) \,.
\end{align}
With this, we find
\begin{align}
\begin{split}
 &2 H (4+3x) - x(1+x) \partial_x H \\
 &= H(5+3x) + \frac16 (1+x) (18 + 3x^2\alpha +2x^3\alpha) > 0 \,,
\end{split}
\end{align}
which completes the proof.

\subsection{Maximum}
\label{sec:approx:maximum}

We first consider the maximum of $a_{\tilde{\varphi}}(r)/\sqrt{a_0 a_b(r)} = \sqrt{M_{\mathrm{eff}}(r)/M_b}$.
The maximum of both $M_{\mathrm{eff}}(r)$ and $\sqrt{M_{\mathrm{eff}}(r)}$ is determined by $M_{\mathrm{eff}}'(r) = 0$.
This condition can be written as
\begin{align}
\label{eq:approx:maxp}
p = \ln(x) - \frac{1}{x} \,,
\end{align}
which can be solved in terms of the Lambert W function,
\begin{align}
 x = \frac{1}{W(e^{-p})} \,.
\end{align}
Using this condition gives further for the maximum of $\sqrt{M_{\mathrm{eff}}/M_b}$ with value $1+\delta$,
\begin{align}
\label{eq:approx:maxx}
\frac{x^2}{6} + \frac{x^3}{9} = \frac{(1+\delta)^2 -1}{\alpha} \,.
\end{align}
This can be solved exactly but the resulting expression is not particularly illuminating.
The size of the right-hand side determines whether the $x^2$ or the $x^3$ term dominates.
For large values, the $x^3$ term dominates, for small values the $x^2$ term dominates.
For large right-hand sides, we have for the maximum
\begin{align}
 x = \left(9 \frac{(1+\delta)^2-1}{\alpha}\right)^{1/3} - \frac12 + \mathcal{O}\left(\left(\frac{\alpha}{(1+\delta)^2-1}\right)^{1/3}\right) \,.
\end{align}
The boundary condition needed to obtain such a maximum is
\begin{align}
 p = \frac13 \ln\left(9 \frac{(1+\delta)^2-1}{\alpha}\right) + \mathcal{O}\left(\left(\frac{\alpha}{(1+\delta)^2-1}\right)^{1/3}\right) \,.
\end{align}

We consider next the maximum of $a_{\mathrm{tot}}/(a_b + \sqrt{a_0 a_b})$, that is, the maximum of the function
\begin{align}
\label{eq:approx:atotfunc}
\frac{\frac{M_{\mathrm{eff}}}{M_b} \frac{1}{x} + \sqrt{\frac{M_{\mathrm{eff}}}{M_b}}}{\frac{1}{x} + 1} \equiv \frac{H/x + \sqrt{H}}{1/x + 1} \,,
\end{align}
assuming $M_{\mathrm{eff}} > 0$.
The maximum of this function is both at a similar location and has a similar value as that of $\sqrt{M_{\mathrm{eff}}}$ -- at least for galaxies.
We first show that the maximum is at a similar location.
To this end, we set the derivative with respect to $x$ to zero, plug in $x + \delta_x$ where $x$ is the maximizer of $\sqrt{M_{\mathrm{eff}}}$, expand to lowest order in $\delta_x$, use $H'(x) = 0$, and solve for $\delta_x$.
This gives
\begin{align}
 \label{eq:approx:deltaxmax}
 \frac{\delta_x}{x} = \left(\frac{2x}{1+x} - \frac{9(1+x)^2(2+\delta)(2+x+2\delta)}{x(3+2x)(1+\delta)^2}\right)^{-1} \,,
\end{align}
where we have additionally used Eq.~\eqref{eq:approx:maxx}, $H(x) = (1+\delta)^2$, $H''(x) = -(1+x)\alpha$, and we have eliminated $\alpha$ by using Eq.~\eqref{eq:approx:maxx}, giving $\alpha = ((1+\delta)^2-1)/ (x^2/6 + x^3/9)$.

The simplest way to understand this expression is to plot it as a function of $x$ for a number of values of $\delta$.
For $\delta \ll 1$, we have that $|\delta_x/x|$ is always smaller than $2.5\%$.
For $\delta = \mathcal{O}(1)$, it can become a bit larger.
For example, up to $\delta = 1.5$, we have that $|\delta_x/x|$ is always smaller than $5\%$.
But even for extremely large values of $\delta$, we have that $|\delta_x/x|$ is always smaller than $14.5\%$.
Thus, the maximum of $a_{\mathrm{tot}}/(a_b + \sqrt{a_0 a_b})$ is indeed at a similar location as that of $a_{\tilde{\varphi}}/\sqrt{a_0 a_b}$.

Now we can evaluate the maximum value of the function Eq.~\eqref{eq:approx:atotfunc}, assuming that the maximizer is close to the one of $\sqrt{M_{\mathrm{eff}}/M_b}$,
\begin{align}
\frac{\frac{M_{\mathrm{eff}}}{M_b} \frac{1}{x} + \sqrt{\frac{M_{\mathrm{eff}}}{M_b}}}{\frac{1}{x} + 1} \approx \frac{(1+\delta)^2 + x (1+\delta)}{1+x} \,.
\end{align}
We can write this as
\begin{align}
 1 + \delta_{\mathrm{tot}} \approx \frac{(1+\delta_{\tilde{\varphi}})^2 + x (1+\delta_{\tilde{\varphi}})}{1+x}
 = (1+\delta_{\tilde{\varphi}}) \left(1 +  \frac{\delta_{\tilde{\varphi}}}{1+x}\right) \,,
\end{align}
where $1 + \delta_{\mathrm{tot}}$ is the maximum value of Eq.~\eqref{eq:approx:atotfunc} and $1 + \delta_{\tilde{\varphi}}$ is the maximum of $a_{\tilde{\varphi}}/\sqrt{a_0 a_b}$.
We consider first small $\delta_{\tilde{\varphi}}$.
Then, the factors relative to MOND are approximately the same in both cases, that is, $1+\delta_{\mathrm{tot}} \approx 1+\delta_{\tilde{\varphi}}$.
(But we note: $\delta_{\mathrm{tot}}$ and $\delta_{\varphi}$ themselves might differ by up to a factor 2 for small $x$).
When $\delta_{\tilde{\varphi}}$ is not small, $x$ is very large, see Eq.~\eqref{eq:approx:maxx}.
At least for galaxies, which have $\alpha \ll 1$.
Thus, the factors relative to MOND are again approximately the same in both cases -- unless $\delta_{\tilde{\varphi}}$ is very large.
But for such large $\delta$, our approximation is anyway no longer valid, see Appendix~\ref{sec:approx:validity}.

\subsection{Optimal boundary condition}
\label{sec:approx:optimal-bc}

We consider the ratio
\begin{align}
 \frac{a_{\tilde{\varphi}}}{\sqrt{a_0 a_b}} = \sqrt{\frac{M_{\mathrm{eff}}}{M_b}} \,,
\end{align}
and require that it stays between $1+\delta$ and $1-\delta$ for some positive $\delta$.
One may wonder what the boundary condition is that allows this requirement to be fulfilled up to the maximum possible radius $r_{\mathrm{max}}$.
We refer to this as the optimal boundary condition.
Here, we show that it is the boundary condition for which $\sqrt{M_{\mathrm{eff}}/M_b}$ has the value $1+\delta$ at its maximum, which we discussed in Appendix~\ref{sec:approx:maximum}.

To this end, we consider this solution for which $\sqrt{M_{\mathrm{eff}}/M_b}$ has the maximum value $1+\delta$ at some radius $r_1$.
This solution will have the value $1-\delta$ at some radius $r_2 > r_1$ since $M_{\mathrm{eff}}$ drops to zero after the maximum.
We refer to this solution's boundary condition $p$ as $p_1$.

Here, we consider $M_{\mathrm{eff}} > 0$.
When $M_{\mathrm{eff}} < 0$, the gravitational force in the AeST model points into the opposite direction from that in MOND.
We are not interested in this regime here.
Thus, we assume $\delta < 1$ which implies $M_{\mathrm{eff}} > 0$.

Any solution with a boundary condition $p$ larger than $p_1$ has a larger value $\sqrt{M_{\mathrm{eff}}}$ everywhere (see Appendix~\ref{sec:approx:increasing-bc}).
This implies that such solutions must have $\sqrt{M_{\mathrm{eff}}/M_b} > 1+\delta$ at $r = r_1$.
Thus, boundary conditions $p$ larger than $p_1$ cannot be optimal.

We then consider solutions with boundary condition $p$ smaller than $p_1$.
Such solutions have a smaller value $\sqrt{M_{\mathrm{eff}}}$ and slope $\partial_r \sqrt{M_{\mathrm{eff}}}$ everywhere (see Appendix~\ref{sec:approx:increasing-bc}).
Thus, these solutions reach their maximum earlier than $r_1$ and this maximum has a value smaller than $1+\delta$.
Starting at this lower maximum, such solutions then go to zero faster (since their slope is more negative).
Thus, these solutions reach $\sqrt{M_{\mathrm{eff}}/M_b} = 1 - \delta$ earlier than $r_2$.
Thus, smaller boundary conditions also cannot be optimal.

It follows that the optimal boundary condition is that which gives $\sqrt{M_{\mathrm{eff}}/M_b} = 1 + \delta$ at the maximum of $M_{\mathrm{eff}}$.
That is, we have $p = p_1$.

The preceding argument implicitly considered negative condensate densities by considering radii beyond the radius where $M_{\mathrm{eff}}$ is maximal.
One may wonder whether anything changes when restricting solutions to have positive densities.
But this is not the case, at least as long as we just cut off the solutions when the density reaches zero and do not continue them with something else (such as a Navarro-Frenk-White halo).
The above argument can be straightforwardly adapted to this case.

Instead of $a_{\tilde{\varphi}}/\sqrt{a_0 a_b}$ we could also consider the total acceleration relative to MOND.
That is,
\begin{align}
 \frac{a_{\tilde{\varphi}} + a_{\tilde{\Phi}}}{a_b + \sqrt{a_0 a_b}} = \frac{\frac{M_{\mathrm{eff}}}{M_b} \frac1x + \sqrt{\frac{M_{\mathrm{eff}}}{M_b}}}{\frac1x + 1} \,.
\end{align}
The argument above can be adapted to this case as well.
That is, the optimal boundary boundary condition is that for which this total acceleration ratio has the maximum value $1+\delta$.

The optimal boundary conditions for $a_{\tilde{\varphi}}$ are, in general, not the same as those for the total acceleration $a_{\tilde{\varphi}} + a_{\tilde{\Phi}}$.
However, in practice, they are very similar and give very similar deviations from MOND at very similar maximum radius.
This follows from the properties of the maximum discussed in Appendix~\ref{sec:approx:maximum}.

\subsection{Relation between $r_{\mathrm{max}}$ and $r_{\mathrm{maxratio}}$}
\label{sec:approx:rmax-vs-rmaxratio}

Here, we consider a solution that is optimal for some $\delta$.
As discussed above and as illustrated in Fig.~\ref{fig:illustrate-bestIC-terminology}, the radius $r_{\mathrm{maxratio}}$ is where the ratio $a_{\tilde{\varphi}}/\sqrt{a_0 a_b}$ has its maximum value, namely $1+\delta$, and $r_{\mathrm{max}}$ is the radius where this ratio drops to $1-\delta$.
Here, we consider the relation between these two radii.
We argue that, usually, $r_{\mathrm{max}}$ is larger than $r_{\mathrm{maxratio}}$ by a fixed factor $1.53$ for small $\delta$.

Based on numerical examples, we first guess a factor around $1.5$.
Then, we calculate first-order corrections to this factor $1.5$ and find that corrections are, in a certain regime, small.
This confirms the validity of our initial guess.
In addition, we find that the first-order corrections are universal and give a factor $1.53$.

To find $r_{\mathrm{max}}$, we must solve $\sqrt{M_{\mathrm{eff}}/M_b} = 1-\delta$.
That is,
\begin{align}
1 + \alpha x^2 \left(\frac12 + \frac19 x(3p + 1 -3 \ln(x))\right) = (1-\delta)^2 \,.
\end{align}
where $x = r_{\mathrm{max}}/r_{\mathrm{MOND}}$.
We write
\begin{align}
 x = f x_0 (1+\epsilon) \,,
\end{align}
where $x_0$ is the radius where this equation is satisfied with $(1+\delta)$ instead of $(1-\delta)$ on the right-hand side, that is to say it is the radius where $M_{\mathrm{eff}}$ is maximal, that is, $x_0 = r_{\mathrm{maxratio}}/r_{\mathrm{MOND}}$.
We later set $f = 3/2$ and let $\epsilon$ parametrize deviations from a factor $1.5$ between $r_{\mathrm{max}}$ and $r_{\mathrm{maxratio}}$, that is, later we find $f(1+\epsilon) \approx 1.53$.

We plug this expression for $x$ back into the equation determining $x$, expand in $\epsilon$, and solve for $\epsilon$.
Then, we eliminate first $p$ and then $\alpha$ using the two conditions we know from $x_0$ being a maximum, namely Eq.~\eqref{eq:approx:maxp} and Eq.~\eqref{eq:approx:maxx}.
This gives
\begin{align}\epsilon = - \frac13 + \frac{
  (2-\delta)(3+2x_0) + 3 f^2(2+\delta) + 2 f^3 x_0 (2+\delta)
}{
  18 f^2 (2+\delta)(f-1 + f x_0 \ln(f))
}\,.\end{align}
We now consider $f=3/2$ and check whether or not $\epsilon$ is actually small.
For large $x_0$, we find
\begin{align}
\epsilon_{f=3/2} = \frac{70 + 19\delta -81(2+\delta)\ln(3/2)}{243 (2+\delta) \ln(3/2)} + \mathcal{O}(1/x_0)\,.
\end{align}

The maximum radius $x_0$ is large as long as $\delta/\alpha \gg 1$, see Appendix~\ref{sec:approx:maximum}.
For galaxies, we typically have $\alpha \lesssim 10^{-4}$ (see Appendix~\ref{sec:optimal}).
Thus, this condition is fulfilled unless $\delta$ is extremely large.
For such large $\delta$, the first order analytical approximation for $M_{\mathrm{eff}}$ is anyway no longer valid, as we show in Appendix~\ref{sec:approx:validity}.
Thus, for galaxies, the large-$x_0$ result is typically valid.

If, in addition, we have $\delta \ll 1$, we find
\begin{align}
\epsilon_{f=3/2} = -\frac13 + \frac{35}{243 \ln(3/2)} + \mathcal{O}(\delta) + \mathcal{O}(1/x_0) \,.
\end{align}
Thus,
\begin{align}
\frac{r_{\mathrm{max}}}{r_{\mathrm{maxratio}}} = \frac32 (1 + \epsilon_{f=3/2}) \approx 1 + \frac{35}{162 \ln(3/2)} \approx 1.53 \,.
\end{align}

\subsection{Validity of approximation}
\label{sec:approx:validity}

We now consider the validity of our first order approximation $M_{\mathrm{eff},1}$ for $M_{\mathrm{eff}}$ from Eq.~\eqref{eq:Meffx}.
Figure~\ref{fig:illustrate-bcdependence} and other plots suggest that it is often a reasonable approximation but also that it becomes worse for larger deviations from MOND $\delta$.
Here, we estimate analytically when this approximation is valid.

To this end, we calculate the 2nd order approximation to $M_{\mathrm{eff}}$ and check when the first and second orders deviate from each other.
To facilitate a quantitative analytical estimate, we do not calculate the 2nd order approximation exactly.
We make additional approximations and assumptions.
The result should still be a reasonable estimate of when the first order approximation is valid.

The 2nd order approximation is obtained by plugging the first order approximation into the right-hand side of Eq.~\eqref{eq:Meffrecursion},
\begin{multline}
 \frac{M_{\mathrm{eff,2}}(x)}{M_b} = \frac{M_{\mathrm{eff},1}(x)}{M_b} - \alpha \int_0^x dx' x'^2
 \int_0^{x'} dx'' \left[
 \frac{1}{x''^2} \left(\frac{M_{\mathrm{eff},1}(x'')}{M_b} - 1\right)
 \right. \\ \left.
 + \frac{1}{x''} \left(s_{\mathrm{eff},1}(x'') \sqrt{\frac{|M_{\mathrm{eff},1}(x'')|}{M_b}}-1\right)
 \right] \,,
\end{multline}
where the subscripts 1 and 2 refer to the first and second order approximations, respectively.

The double integrals in this expression can be done analytically for the part linear in $M_{\mathrm{eff},1}$.
We find
\begin{multline}
\frac{M_{\mathrm{eff},2}^{\mathrm{lin}}(x)}{M_b} = \frac{M_{\mathrm{eff},1}(x)}{M_b} \\- \frac{x^4 \alpha^2}{1800} \left(225 + (62+60p)x - 60x \ln(x) \right) \,.
\end{multline}
Below we express radii relative to where $M_{\mathrm{eff},1}$ has its maximum,
\begin{align}
 \hat{x} \equiv \frac{r}{r_{\mathrm{maxratio}}} \equiv \frac{x}{x_{mr}} \,.
\end{align}
We can also eliminate $p$ by using the relation $p = \ln(x_{\mathrm{mr}}) - 1/x_{\mathrm{mr}}$.
This gives
\begin{multline}
\frac{M_{\mathrm{eff}}^{\mathrm{lin}}(\hat{x})}{M_b} \approx \frac{M_{\mathrm{eff},1}(\hat{x})}{M_b} \\- \frac{\hat{x}^4 x_{\mathrm{mr}}^4 \alpha^2}{1800} \left(225 - 60 \hat{x} + 62 x_{\mathrm{mr}} \hat{x}\left(1 - \frac{60}{62} \ln(\hat{x}) \right) \right) \,.
\end{multline}
We consider $\hat{x}$ between $0$ and $1$.
The function $\hat{x} (1 - 60/62\ln(\hat{x}))$ grows from $0$ to $1$ in this $\hat{x}$ interval.
Thus, when $x_{\mathrm{mr}}$ is large, that is, when $r_{\mathrm{maxratio}}$ is large compared to $r_{\mathrm{MOND}}$, this term will dominate, except at very small radii $\hat{x}$.
Otherwise, the $225- 60\hat{x}$ term dominates.

We now consider the remaining terms in $M_{\mathrm{eff},2}/M_b$, that is, those with $\sqrt{M_{\mathrm{eff},1}/M_b}$.
Here, the double integrals are not straightforward to do analytically.
Therefore, we make an additional approximation and we restrict ourselves to radii $r$ before $M_{\mathrm{eff},1}$ reaches its maximum, that is, we only consider $r \leq r_{\mathrm{maxratio}}$.
Concretely, we assume that $\sqrt{M_{\mathrm{eff},1}/M_b} = a_{\tilde{\varphi},1}/\sqrt{a_0 a_b}$ grows linearly from $1$  at $r=0$ to its maximum $1+\delta$ at $r = r_{\mathrm{maxratio}}$,
\begin{align}
 \frac{a_{\tilde{\varphi},1}(r)}{\sqrt{a_0 a_b(r)}} = \sqrt{\frac{M_{\mathrm{eff},1}(r)}{M_b}} = 1 + \delta \frac{r}{r_{\mathrm{maxratio}}} \,, \quad r \leq r_{\mathrm{maxratio}} \,.
\end{align}
From Fig.~\ref{fig:illustrate-bestIC-terminology} we see that this underestimates $M_{\mathrm{eff},1}$ at small radii and then overestimates it at larger radii toward $r_{\mathrm{maxratio}}$.
This is a rough approximation but we have verified numerically that it captures the validity of the first order approximation reasonably well.
With this, we have
\begin{align}
 \frac{M_{\mathrm{eff},2}^{\mathrm{sqrt}}(x)}{M_b} = - \frac{\alpha \delta}{4} x_{\mathrm{mr}}^3 \hat{x}^4 \,.
\end{align}

The remaining task is to sum $M_{\mathrm{eff},2}^{\mathrm{lin}}$ and $M_{\mathrm{eff},2}^{\mathrm{sqrt}}$ and determine when this sum deviates from $M_{\mathrm{eff},1}$.
More specifically, we are interested mostly in $a_{\tilde{\varphi}}$ so we are interested in the square root.
We parametrize the deviation by a parameter $q$,
\begin{align}
 \label{eq:approx:validityq}
 \sqrt{\frac{M_{\mathrm{eff},2}(\hat{x})}{M_{\mathrm{eff},1}(\hat{x})}} \stackrel{!}{=} 1 - q \,.
\end{align}
This relation determines the radius $\hat{x}$ where the 2nd order becomes important if we want to trust the first order approximation up to a fraction $q$.

Unfortunately, this equation is not straightforward to solve analytically.
Thus, we make further approximations.
First, we use the linear approximation $\sqrt{M_{\mathrm{eff},1}/M_b} = 1 + \delta \hat{x}$ also in  the denominator of Eq.~\eqref{eq:approx:validityq} and the first term in $M_{\mathrm{eff},2}^{\mathrm{lin}}$.
Second, we neglect the $\hat{x}(1 - (60/62) \ln(\hat{x}))$ term in $M_{\mathrm{eff},2}^{\mathrm{lin}}$.

The reasoning behind the second of these approximations is the following.
As argued above, the $\hat{x}(1 - (60/62) \ln(\hat{x}))$ term is important only for large $x_{\mathrm{mr}}$.
But in this case, the double-integral contributions to $M_{\mathrm{eff}}^{\mathrm{lin}}$ are anyway small compared to $M_{\mathrm{eff}}^{\mathrm{sqrt}}$ so it does not matter.
Indeed,
\begin{align}
 \left.\frac{M_{\mathrm{eff},2}^{\mathrm{lin}} - M_{\mathrm{eff},1}}{M_{\mathrm{eff},2}^{\mathrm{sqrt}}}\right|_{x_{\mathrm{mr}} \gg1} \approx \frac{31}{25} \frac{1}{x_{\mathrm{mr}}} \frac{(1+\delta)^2-1}{\delta} \hat{x} \left(1 - \frac{60}{62} \ln(\hat{x})\right) \,,
\end{align}
where we used $\alpha = 9 ((1+\delta)^2-1)/x_{\mathrm{mr}}^3$ which is valid for $x_{\mathrm{mr}} \gg 1$, see Eq.~\eqref{eq:approx:maxx}.
Since $\hat{x} \leq 1$ and $x_{\mathrm{mr}} \gg 1$, this ratio is small unless $\delta \gtrsim x_{\mathrm{mr}}$.
Thus, for our purposes this ratio is small.
We are not interested in very large deviations from MOND $\delta$ which would be required for $\delta \gtrsim x_{\mathrm{mr}} \gg 1$.

Equation~\eqref{eq:approx:validityq} then becomes
\begin{align}
 \frac{- \frac{\alpha \delta}{4} x_{\mathrm{mr}}^3 \hat{x}^4 - \frac{\alpha^2}{8} \hat{x}^4 x_{\mathrm{mr}}^4 \left(1 - \frac{4}{15} \hat{x}\right)}{(1+\delta \hat{x})^2} \stackrel{!}{=} (1-q)^2-1 \,.
\end{align}
This is still not solvable analytically.
Thus, our final approximation will be to neglect the $-4/15 \hat{x}$ term.
This is again a rough approximation but, as mentioned above, we have verified numerically that the result is reasonable.
Thus, we need to solve
\begin{align}
\sqrt{\frac14 \alpha x_{\mathrm{mr}}^3 \delta + \frac18 \alpha^2 x_{\mathrm{mr}}^4} \frac{\hat{x}^2}{1 + \delta \hat{x}} \stackrel{!}{=} \sqrt{1 - (1-q)^2} \,.
\end{align}

We finally find for the radius $r_{\mathrm{approx}}$ up to which the first order approximation is valid to within a fraction $q$,
\begin{align}
 \label{eq:validity}
 \frac{r_{\mathrm{approx}}}{r_{\mathrm{maxratio}}} = \frac12 \left(\beta \delta + \sqrt{(\beta \delta)^2 + 4 \beta}\right) \,,
\end{align}
where
\begin{align}
\beta = \frac{\sqrt{2}}{3} \sqrt{\frac{1-(1-q)^2}{\delta^2 (2+\delta)}} \sqrt{\frac{
  (3+2x_{\mathrm{mr}})^2
}{
  (3 + 2x_{\mathrm{mr}})x_{\mathrm{mr}} + 9 (2+\delta)
}}\,,
\end{align}
where we used Eq.~\eqref{eq:approx:maxx} to eliminate $\alpha$.

We note that we assumed $r \leq r_{\mathrm{maxratio}}$ when deriving this estimate.
Thus, if our expression for $r_{\mathrm{approx}}/r_{\mathrm{maxratio}}$ is larger than $1$, the first order approximation is valid up to at least $r_{\mathrm{maxratio}}$ but the precise value of $r_{\mathrm{approx}}/r_{\mathrm{maxratio}}$ is not meaningful.

The most important implication of our estimate is that the first order approximation is good for small $\delta$ (except for extremely restrictive values of $q$).
For larger $\delta$, the approximation is worse.
Explicitly, for small $\delta$ the quantity $\beta$ scales as
\begin{align}
\beta_{\delta \ll 1} \sim \sqrt{q}/\delta
\end{align}
so that $r_{\mathrm{approx}}/r_{\mathrm{maxratio}}$ scales as
\begin{align}
\left.\frac{r_{\mathrm{approx}}}{r_{\mathrm{maxratio}}}\right|_{\delta \ll 1} \sim \sqrt{\beta} \sim \left(\frac{q}{\delta^2}\right)^{1/4} \,.
\end{align}
When $\delta$ is small, this ratio is large, so the first order approximation is valid.
If we demand a better approximation (i.e., a smaller $q$), the radius $r_{\mathrm{approx}}$ becomes smaller.

In the opposite limit, $\delta \gg 1$,
\begin{align}
\beta_{\delta \gg 1} \sim \sqrt{q/\delta^{3}}\;\mathrm{or}\;\sqrt{q/\delta^{4}} \,,
\end{align}
where the power $3$ applies for sufficiently large $x_{\mathrm{mr}}$ and the power $4$ for sufficiently small $x_{\mathrm{mr}}$.
In both cases does $r_{\mathrm{approx}}/r_{\mathrm{maxratio}}$ tend to zero for large $\delta$.
So the first order approximation is not valid much.

\end{appendix}

\end{document}